\documentclass[sigconf,authorversion,nonacm]{acmart}
\usepackage{savesym}

\usepackage{graphicx}
\usepackage{subcaption}
\usepackage{balance}
 \usepackage{enumitem}

\newbool{finalversion}
\setbool{finalversion}{true}

\AtBeginDocument{%
  }

\acmConference[]{}{}{}
\acmPrice{15.00}
\acmISBN{978-1-4503-XXXX-X/18/06}


\makeatletter
\def\@copyrightspace{\relax}
\makeatother

\begin{document}

\title[Imagine a dragon]{Imagine a dragon made of seaweed: How images enhance learning in Wikipedia}

\author{Anita Silva$^\spadesuit$, Maria Tracy$^\spadesuit$, Katharina Reinecke$^\spadesuit$, Eytan Adar$^\diamondsuit$, Miriam Redi$^\clubsuit$}

\affiliation{
 \institution{$\spadesuit$ Paul G. Allen School of Computer Science \& Engineering, University of Washington; $\diamondsuit$ School of Information,  University of Michigan; $\clubsuit$Wikimedia Foundation}
}

\renewcommand{\shortauthors}{Silva et al.}

\begin{abstract}
Though images are ubiquitous across Wikipedia, it is not obvious that the image choices optimally support learning. When well selected, images can enhance learning by dual coding, complementing, or supporting articles. When chosen poorly, images can mislead, distract, and confuse. We developed a large dataset containing 470 questions \& answers to 94 Wikipedia articles with images on a wide range of topics. Through an online experiment (n=704), we determined whether the images displayed alongside the text of the article are effective in helping readers understand and learn. For certain tasks, such as learning to identify targets visually (e.g., \textit{which of these pictures is a ``gujia''}?)\footnote{For the curious: Gujia are sweet moon-shaped dumplings native to the Indian subcontinent. }, article images significantly improve accuracy. Images did not significantly improve general knowledge questions (e.g., \textit{where are gujia from?}). Most interestingly, only some images helped with visual knowledge questions (e.g., \textit{what shape is a gujia?}). Using our findings, we reflect on the implications for editors and tools to support image selection.

\end{abstract}


\maketitle
\section{Introduction}

\begin{figure*}[!t]
\centering
\begin{subfigure}[b]{0.38\textwidth}
  \includegraphics[width=\linewidth]{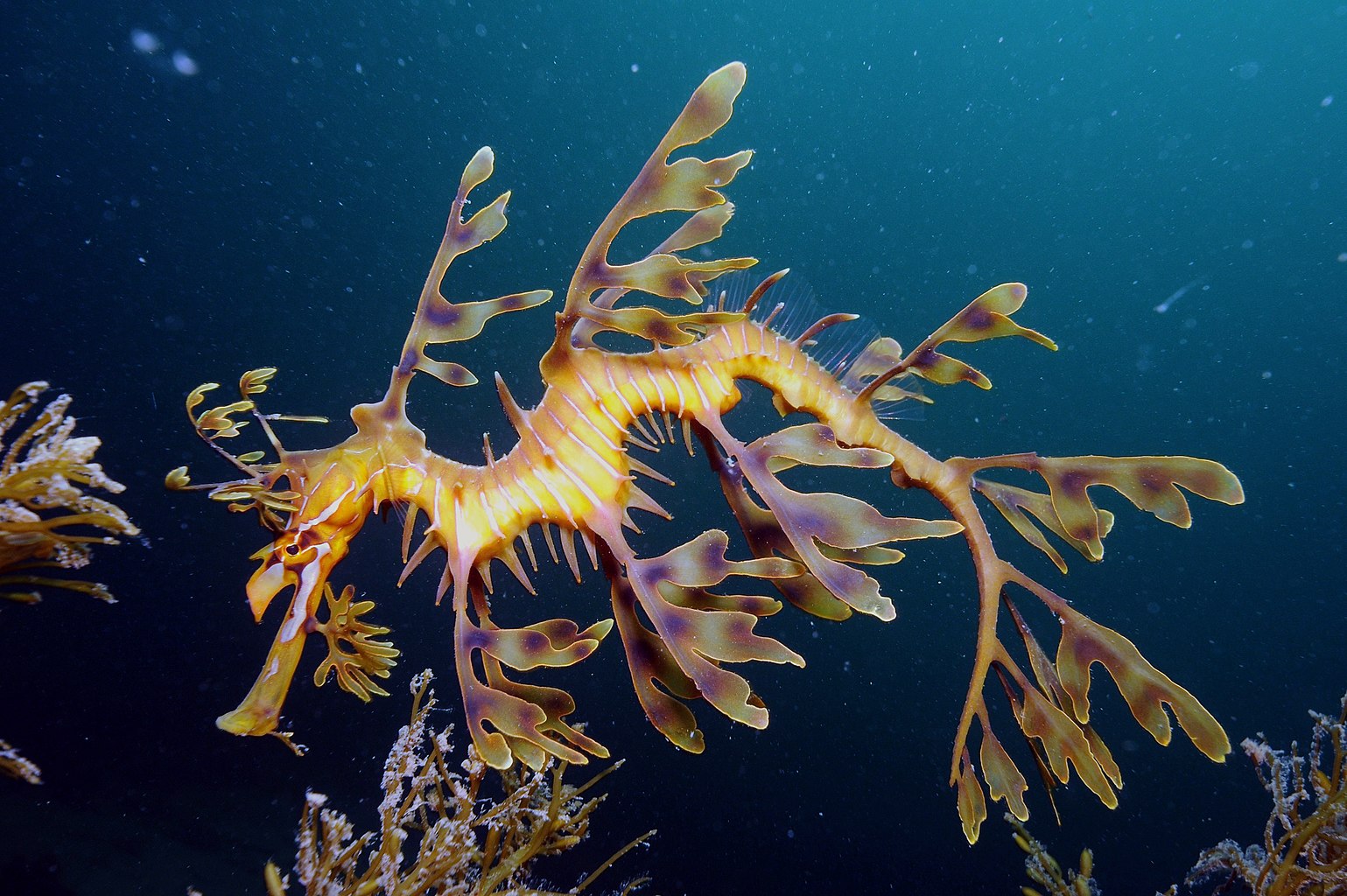}
  \caption{}
   \end{subfigure}
   \hspace{1cm}
\begin{subfigure}[b]{0.43\textwidth}
  \includegraphics[width=\linewidth]{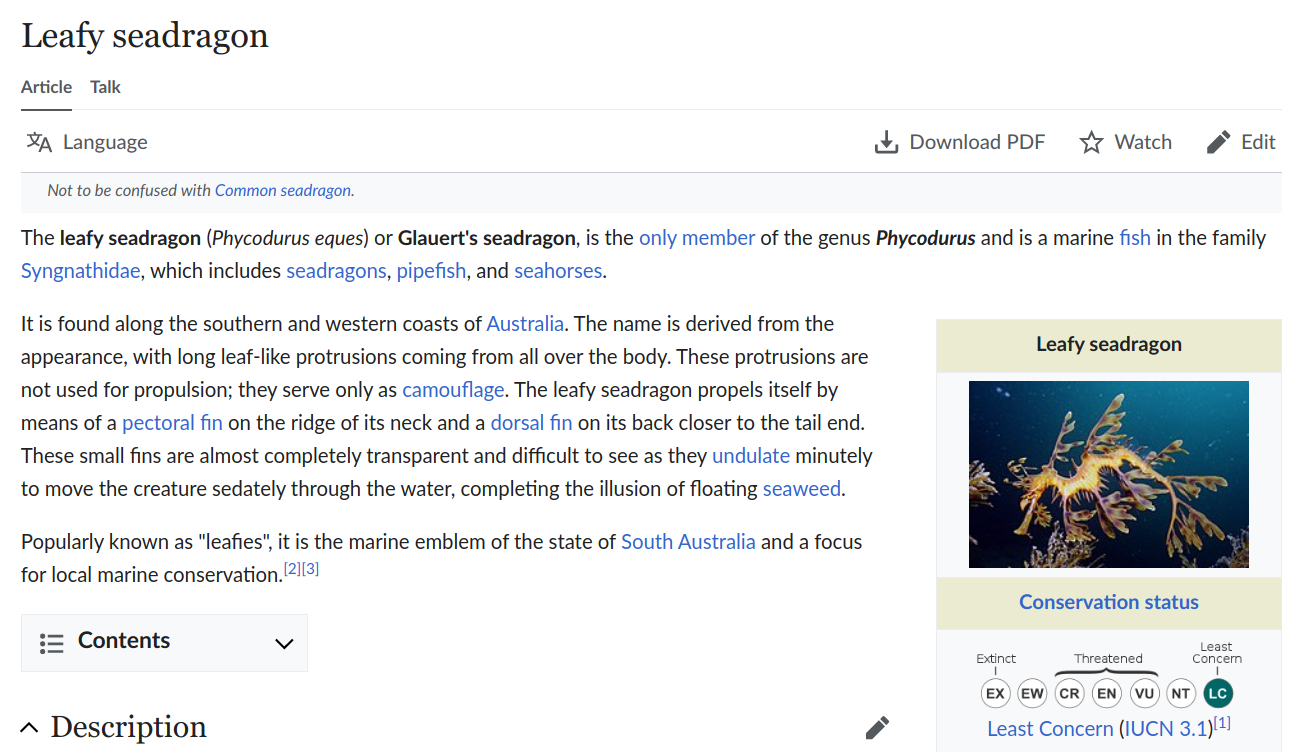}
  \caption{}
   \end{subfigure}
\hfill
  \caption{(a) An image of a \textit{leafy seadragon} (source: \url{https://commons.wikimedia.org/wiki/File:Leafy_Seadragon_on_Kangaroo_Island.jpg}) used at the top of the page (within the infobox) for the English Wikipedia page on the topic (b). This will be the first image readers encounter on the page.}
  \label{fig:leafyseadragon}
\end{figure*}

One of the main motivations for visiting Wikipedia is intrinsic learning~\cite{singer2017we}. This is due in large part to the significant editorial attention that curates Wikipedia content toward a clear, factual knowledge repository. While most attention is focused on textual content, Wikipedia sites are a multimedia experience. Images complement and augment millions of pages. For example, the reader visiting the Leafy Seadragon page may find it difficult to visualize how, ``the lobes of skin that grow on the leafy seadragon provide camouflage, giving it the appearance of seaweed.''~\cite{seadragon}.  When a Wikipedia editor introduces a high-quality image (see Figure~\ref{fig:leafyseadragon}), that image can significantly illuminate the subject. 

To support the use of such images, Wikimedia allows for freely licensed multimedia files to be uploaded to the Wikimedia Commons site. As of February 2024, more than 103 million documents---largely photos, maps, and diagrams---have been uploaded with the central purpose of providing educational benefits (i.e., ``providing knowledge; instructional or informative.''~\cite{purposepage}). In the context of Wikipedia, images can be curated, added, removed, modified, placed into galleries, captioned, referenced, and otherwise editorialized. 

Although there is significant research on what makes for good text on Wikipedia and how editors craft it (e.g.,~\cite{adler08,kittur08,wilkinson07}), there is, unfortunately, less data on image use. This gap is important as good images can enhance learning (e.g., through dual coding~\cite{paivio1990mental}), while `bad' images can distract and confuse (e.g., through seductive details~\cite{sundararajan2020keep} or structure interference~\cite{SCHNOTZ2003141}).  
With a for-profit textbook or encyclopedia, authors or editors may be able to find the ideal image. In the case of Wikipedia, it is not clear that the ``best'' representation of a concept can as easily be acquired or created. Wikipedia presents an added complexity, as images and associated text are manually curated by a distributed community of editors and must generally comply with free-licensing rules. While the Wikimedia Commons image collection is vast, it is still bounded by the availability of free content and by the scale of volunteer efforts. The consequence is that we do not know if and when Wikipedia-embedded images can help in learning. 

Images 
can serve in multiple roles for multiple kinds of learning (see~\cite{Marsh2003:0022-0418:647} for a survey). Critically, images on Wikipedia are not isolated---they are embedded in the context of text. Thus, an image can function in a way that closely relates to that text (e.g., to exemplify a concept) but may also have little relation to it (e.g., a decorative purpose) or go beyond it (e.g., interpret). In the case of Wikipedia, decoration roles are unlikely, as there is a strong emphasis on the use of images for educational purposes~\cite{purposepage}. Similarly, interpretive roles should be avoided as they may go against Wikipedia's `No original research' policy~\cite{noresearch}. Although this leaves most Wikipedia images squarely in the `close relation to text' category, these images can still support different use cases. Maps can help the reader orient themselves or provide context (e.g., where an unfamiliar town is situated). A flow diagram can illustrate complex mechanisms (e.g., how the Kreb's cycle works).

To investigate the role of images in learning on Wikipedia, this paper investigates the following research question: \textbf{Do Wikipedia articles with images help 
answer knowledge comprehension questions accurately compared to articles without an image?} We address this question by designing a dataset and a crowdsourced experiment to (i)~quantify the importance of images for knowledge acquisition on Wikipedia, and (ii) understand which image characteristics make them useful for learning.
We focus on images that are largely illustrative of the Wikipedia concept (in the classification of~\cite{Marsh2003:0022-0418:647} the image `reiterates', or more specifically, `concretizes' or `exemplifies'). For example, the image associated with the Leafy Seadragon article in Figure~\ref{fig:leafyseadragon} provides an illustrative example of what a leafy seadragon looks like. A photograph of Donatello's statue of David makes a concrete illustration of what the statue looks like. We generated a learning assessment dataset from 94 selected articles and corresponding images of this type. In a pre-registered\footnote{\url{https://aspredicted.org/JJQ_GVQ}} online experiment with 704 participants recruited through LabintheWild~\cite{labinthewild} and Prolific, we determined the helpfulness of images integrated into these articles for a number of learning tasks. Specifically, we assessed: (1) \textit{Image Recognition Questions}--can the reader learn to distinguish a new image of a target concept (e.g., is this a leafy seadragon)?; (2) \textit{Visual Knowledge Questions}--can the reader learn visual properties of the concept (e.g., what color are leafy seadragons)?; and (3) \textit{General Knowledge Questions}--can the reader learn general knowledge information about the concept (e.g., where are leafy seadragons found)? In all, we crafted 470 questions across all three types. These, of course, assess only a small portion of possible learning objectives. From the perspective of learning, we target the lowest levels of Bloom's learning objective taxonomy--Recall or Recognize~\cite{anderson2001taxonomy,bloom1956taxonomy}. Nonetheless, these objectives are consistent with those of the majority of Wikipedia visitors in getting an overview or fact-checking~\cite{lemmerich2019world,singer2017we}.

Our results show that images can support learning, but only for some tasks. In our study, images significantly improved Image Recognition tasks but did not improve General Knowledge learning. However, we find that images do not appear to hurt the learning tasks---a concern if readers overly focus on the image and do not engage with the content.  The results for answering Visual Knowledge questions are more mixed. We found that in some cases images can be consistent or inconsistent with textual content. For example, the text may indicate that a specific gem's coloring is ``blue or green.'' However, the image may only show a blue example. When asked, ``can the gem be green?'' the reader may not be able to correctly answer. With these \textit{inconsistent} images, a reader would not answer the question correctly if they were only focused on the image and ignored or did not read the text. In contrast, with \textit{consistent} images, the answer is apparent from the image regardless of the text. 

We identify several implications of these results, including better support tools for Wikipedia editors, policies, and potential new forms of `wikiwork.' In addition to these empirical results and design implications, we contribute a novel dataset of articles, images, and assessments that can be used as a baseline for evaluating images on Wikipedia\footnote{\url{https://github.com/datamazelab/wikiimages}}.

\section{Related Work}
Literature from different disciplines informed the research in this paper. To contextualize our work, we first discuss findings on the role of images in learning and reading comprehension. Additionally, we place our research within related work on Wikipedia as a learning platform and the impact of various design choices made in the presentation of content on Wikipedia.

\subsection{The role of text illustrations}
Several projects from experimental psychology have focused on the role of images in learning and reading comprehension.
Research work investigating images' \textit{cognitive} function (as defined by Levie and Lentz \cite{levie1982effects}) to facilitate knowledge acquisition found mixed results. In some cases, images in association with text help support learning \cite{guo2020you}, especially when images are correctly captioned~\cite{bernard1990using}, and when graphics are designed to support the interpretation of textual information~\cite{renkl2017studying}. On the other hand, the added visual complexity can challenge the reader's comprehension \cite{guo2020you}.
In online settings, images have been found to slow reading time without having a significant influence on content retention \cite{beymer2007eye,li2019impact}.

Visual content in conjunction with text can have an \textit{attentional} role, drawing attention to information in textual form, or as an \textit{affective} pull that enhances emotions~\cite{leerobbinsaffective, levie1982effects}.  Images can also have an effect on reader motivation and prejudicial behavior: for example, the presence of images of women in science books has a positive impact on the performance of female students \cite{good2010effects}. 

There are a number of theories and empirical results that support the notion that well-crafted or selected visual images can aid in learning. For example, dual coding theory argues that two different cognitive systems (verbal and imagery/nonverbal) retain different types of information, each with its own advantage~\cite{paivio1990mental}. Because these systems are connected, the theory posits that including images, ``may better support retention of the material as it provides learners with two ways to memorize information''~\cite{vekiri02}. However, certain images may disrupt learning (e.g., seductive images, which contain interesting but irrelevant information~\cite{harp1997role}) . 

Although several alternative theories seek to explain the interaction between text and visual signals, there is consistent empirical evidence that \textit{well-selected/crafted} images can aid in learning~\cite{clark2010graphics}. Our work seeks to broadly validate these results in the context of \textit{crowd-selected} images. 

\subsection{Wikipedia as a learning platform}
One of the main reasons why readers access Wikipedia is intrinsic learning~\cite{singer2017we}, especially in developing or newly industrialized countries \cite{lemmerich2019world}. 
For example, Wikipedia is one of the most visited resources for health information \cite{laurent2009seeking}, a pattern that was particularly evident during the COVID-19 outbreak \cite{covidpost}. Although intrinsic learning may be more common, extrinsic motivation (e.g., grades~\cite{head2010today}) can also drive visits. An analysis of the motivations of Wikipedia users based on publicly available data reveals that readers who use English Wikipedia for intrinsic learning or school tasks are more likely to browse articles about mathematics, biology, and history/philosophy~\cite{wwrwdata}.
Research has shown that Wikipedia is widely used by faculty and students alike
\cite{knight2012wikipedia}, and its integration into instructional settings can have positive effects \cite{wannemacher2010wikipedia}, particularly when teaching practices include article creation \cite{evenstein2017wikipedia}. Education is at the core of the Wikimedia Foundation's mission ``to empower and engage people around the world to collect and develop educational content \ldots and to disseminate it effectively and globally''~\cite{mission}. Understanding how images support both real-world use and the mission of the underlying platform becomes a critical task.

\subsection{Design choices on Wikipedia}
Researchers have investigated various aspects of the behavior of Wikipedia readers. These include topic preferences~\cite{10.1145/2631775.2631805}, engagement with links in different positions \cite{dimitrov2016visual},  how readers perceive latency and system performance on the site~\cite{salutari2019large}, and how they engage with images \cite{rama2022large}, citations~\cite{piccardi2020quantifying} and external links~\cite{piccardi2021value}.
However, while Wikipedia is widely used in instructional settings and applications, only a few research works explored how Wikipedia content and layout support reading comprehension and learning. 

For example, Wikipedia has been used to find prerequisite relations for online learning courses~\cite{sayyadiharikandeh2019finding}, or to test the optimal font size for online reading comprehension~\cite{rello2016make} . However, to our knowledge, no previous work has explicitly tested the role of images in reading and understanding Wikipedia content. 

In cases where images have been the subject of research articles, it has primarily been to analyze the quantity and diversity of Wikipedia images. 
Interdisciplinary research exposed the importance of visual content in quantifying Wikipedia's monetary value \cite{heald2015valuation, erickson2018commons}. Recent work measured the diversity of Wikipedia images across languages~\cite{he2018the_tower_of_babel} and gender gaps \cite{beytia2022visual}.

Few works have investigated the role of images on Wikipedia from a \textit{user} perspective. Viegas et al. focused on understanding how and why editors use images when contributing to Wikipedia and describing the editor communities that curate its visual content ~\cite{viegas2007visual}.
Navarrete et al. \cite{navarrete2020image} found that images of paintings on Wikipedia are used to illustrate generic topics beyond art, reaching a larger pool of readers.
Recently, Rama et al. studied how readers interact with images on English Wikipedia \cite{rama2022large} by analyzing traffic logs from Wikimedia Foundation servers. They found that on average, readers click on images 3 out of 100 times they read Wikipedia pages, and that page previews with images are less likely to be loaded in full. While this work provides an interesting quantification of readers' \textit{engagement} with images, it does not provide any insights as to \textit{why} and \textit{how} readers use visual content when browsing the encyclopedia. In this work, we begin to address these questions by studying the effect of images on knowledge acquisition.

\section{A Dataset for Text and Images}
To test knowledge acquisition, we first built a dataset of multiple choice questions, with corresponding text snippets and images, based on Wikipedia articles. 
Our dataset includes three types of questions, informed by prior work on learning and information seeking in Wikipedia~\cite{head2010today, singer2017we,lemmerich2019world}: 

\begin{itemize}
    \item \textbf{General Knowledge Questions} to test the understanding and recall of generic knowledge. These questions ask about either historical, metric, or other non-visual factual aspects of the article topic. Example: \textit{What culture are the Stone spheres of Costa Rica generally attributed to?} %

    \item \textbf{Visual Knowledge Questions}, to measure understanding and recall of the visual characteristics of the subject of the article as described in the article text. These questions are often about color, material, structure, design, and other aspects that visually describe the subject of the article. Example: \textit{What are most of the Stone spheres of Costa Rica made of?} 

    \item \textbf{Image Recognition Questions}, to measure our visual understanding and recall of a topic. These questions test participants' ability to identify the image that best depicts an article topic. Example: \textit{Which of these images correspond to the Stone Spheres of Costa Rica?}
\end{itemize}

To create these questions, we followed a semi-automated procedure that ultimately yielded 470 questions from 94 articles.

\subsection{Article Selection}
We started by creating a controlled subset of relevant English Wikipedia articles. We used an automated process to identify broad topical categories of general interest. 
We created our dataset from a July 2019 snapshot of the English Wikipedia and selected articles classified as Food and Drink, Arts, Architecture, or Biology (e.g., birds, minerals, etc.), using the language-agnostic topic model developed by Johnson et al, \cite{johnson2021language}. 
Of these, only articles with a `lead' image were retained (an image in the infobox or main section of the article\cite{manualofstyle}). Small icons, such as logos or flags, were ignored.

We then selected specific articles that were less well-known, since articles that were too familiar to the broad population would likely lead to questions that were too easy (potentially creating a high baseline score).
To identify `less known' articles, we sub-selected pages that had under 6000 views per month (9th decile) and were less than 15k characters (removing any overly detailed articles). These two requirements ensured that the articles were less popular from the viewers' and editors' perspectives. We treated this as a proxy for articles that were less likely to be familiar to an `average' reader.

When creating both visual and general knowledge questions, we wanted to ensure that the article text provided sufficient information to answer them. For example, the article text would need to mention that Donatello’s David is in Italy and Detroit-style pizza is traditionally rectangular. In piloting question generation, we found that most information for visual knowledge questions came from specific sub-sections of the articles. Specifically, these sections were: \textit{Description, Characteristics, Architecture, Appearance, Design, Details, Morphology, Anatomy and physiology, Construction, Size(s), and Physiology}. Although other sections may also contain visual descriptions, we found through inspection that these covered most of the cases. Thus, we retained articles that had at least one of these sections. In the end, we generated a list of 31k articles. Note that these were simply used as a seed set for our annotators. The articles roughly conformed to our views and length requirements and had a high likelihood of being usable for question creation.

\subsection{Question Generation}\label{sec:q_generation}
While we originally experimented with using Mechanical Turk for this process, we found that the questions produced were unsatisfactory. They were deemed too easy, convoluted, having non-related distractors, or poor images for the image recognition question type. This is consistent with the observation that good assessments are nuanced and may be difficult to create~\cite{haladyna2004developing}. Although the questions produced by the crowd workers were largely unusable, 20 (of 60) articles that they selected were evaluated by the research team as appropriate and used in the experiment. Hence 20 out of the 94 articles used in this study were produced through the Mechanical Turk pilot study.

To build up the remaining set, six members of the research team manually picked pages from the automatically-generated list of articles. Articles that discussed abstract or extremely specialized ideas were not selected for this study due to their conceptual difficulty. The research team then created five questions per article: two general knowledge questions, two visual knowledge questions, and one image recognition question, for a total of 470 questions across 94 articles. For each question, we created a set of multiple choice answers, containing the correct answer, and up to three distractor answers. In addition, we specified the titles of the sections in the articles that contain the answers to the questions.

The question generation process was highly iterative with the research team initially working together and discussing article selection, potential questions, and answers. After these initial discussions, each team member picked a set of 5-10 articles and generated a set of questions and answers. These were subsequently reviewed by other members of the team.

\paragraph{Guidelines for Question and Answers} 
Following the process described above, we developed a set of guidelines to generate a robust set of knowledge acquisition questions. Questions had to be (1) \textit{self-explanatory:}, i.e., not referring to other questions and answerable from memory without still having access to the Wikipedia article seen before; (2) \textit{interesting}, exciting questions that could be part of a trivia game; (3) \textit{specific}, namely not containing ambiguous terms.

Additionally, we developed specific guidelines for choosing general knowledge, visual, and image recognition questions. In the structure of a learning objective framework such as Bloom's~\cite{anderson2001taxonomy,bloom1956taxonomy}) the questions assess to test recall of specific facts (e.g., ``the student will recall who commissioned Donatello's David''). Assessments at this level are more easily structured as multiple response questions~\cite{haladyna2004developing}. Note that while these questions test \textit{comprehension} (and specifically, recall), they were not designed around deeper learning tasks (e.g., summarizing, inference, etc.) or knowledge dimensions (e.g., conceptual, procedural, etc.).

 Visual knowledge questions were created only using article text and did not rely on the article's images. We selected visual knowledge questions in such a way that there was \textit{plausibly} an image that represented \textit{the entity itself} which could be used to answer the question (e.g., a photograph or sketch of Donatello's David). Of course, this \textit{need not be} the image that is actually used on the Wikipedia page. When crafted this way, visual knowledge questions tended to ask about visually salient features (e.g., color), visually comparable features (e.g., size in relation to similar organism), or other physical properties that could be determined with some prior knowledge (e.g, building material). Visual knowledge questions were also tagged with a binary value--consistent or inconsistent--indicating if the question could be answered by only looking at the image for the article (specifically the `lead' or 'main' image in use on Wikipedia). For example, for the article about the Carolina Reaper pepper, the image has three examples of the pepper, but very few cues about scale. Thus, the question ``When fully ripe, about how big is a Carolina Reaper pepper?'' is marked as `inconsistent.' However, the question ``What texture is the Carolina Reaper Pepper?'' is answerable from the image (and thus consistent).
 
General knowledge questions were factual questions that did not have to do with visual properties of the subject and could not be answered by looking at images. These were intended to be simple, factual details, often with one-word answers (e.g., what part of the world or from what country is something? when was something built or created? etc.). Visual and general knowledge questions were structured as multiple choice with a single right answer and up to three distractors. Distractors were generated by members of our research team. Following suggested guidelines~\cite{haladyna2004developing}, the distractors were in the same class as the true answer (e.g., other country names in areas of the world geographically close to the true answer). However, distractors were not intended to create trick questions or confusion.

To create image recognition questions, our guideline suggested finding one picture of the entity that was not in use on the Wikipedia page and a number of distractors (all from Wikimedia Commons). To find an appropriate target image, we suggested a number of candidates, including: lead images used in other language pages, images previously used on the Wikidata item page, or images in the same category in the Wikimedia Commons as the current image. For example, for the Turkish pastry Boyoz, all language editions utilized the same lead image. However, the Commons category page for the image (\url{https://commons.wikimedia.org/wiki/Category:Boyoz}) provided six alternatives. Distractor images were found by looking at related or higher-level categories to the entity. For example, the page for Boyoz links to \url{https://en.wikipedia.org/wiki/List_of_pastries}. From here we see other stuffed pastry types (e.g., the Scottish Bridie), which have a related but distinct appearance. All generated questions--general, visual, image recognition--were later validated for `baseline familiarity' (i.e., ease), as described below.

\begin{figure*}[t]
\centering
\begin{subfigure}[b]{0.4\textwidth}
  \includegraphics[width=\linewidth]{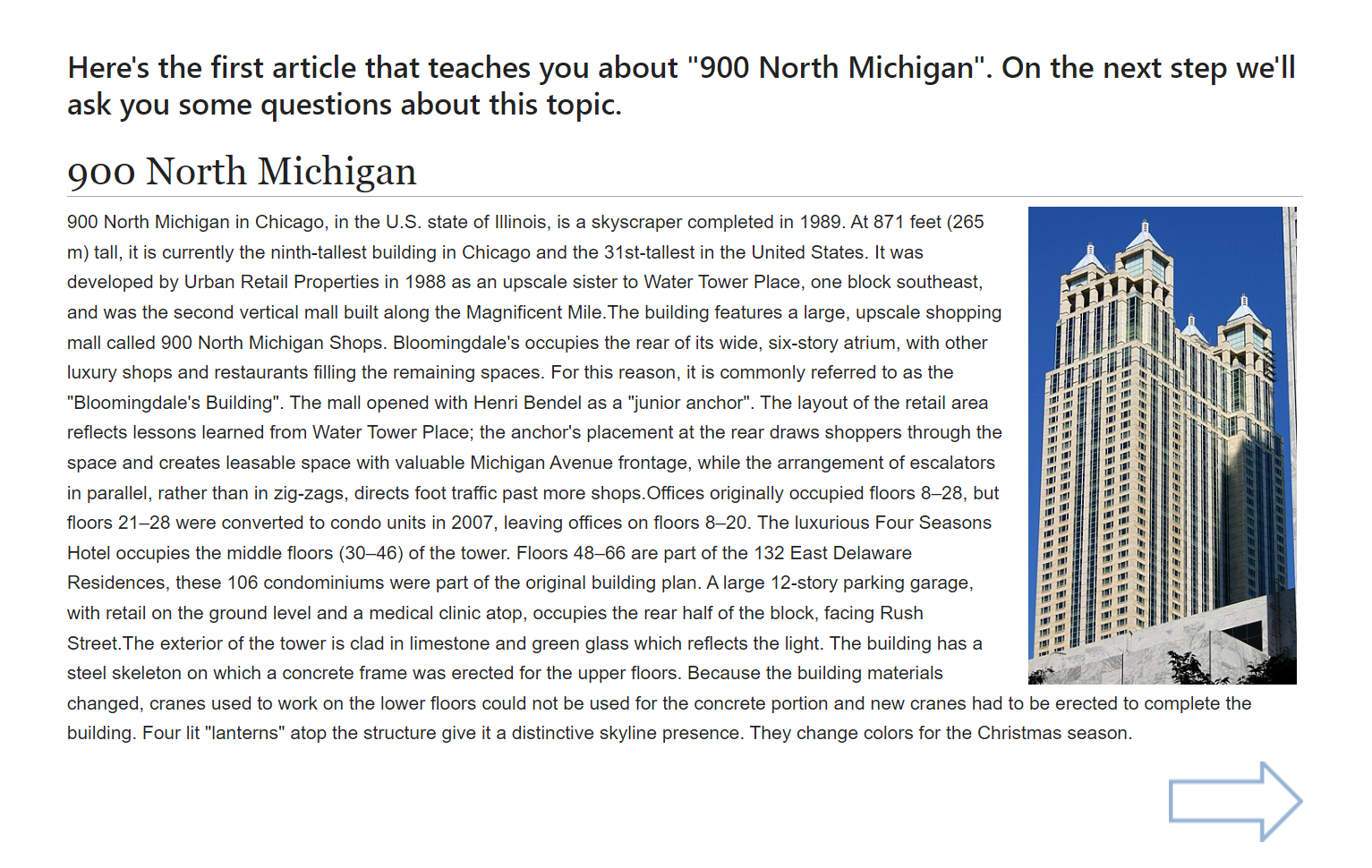}
  \caption{Article displayed with an image}\label{fig:article_w_img}
\end{subfigure}
\begin{subfigure}[b]{0.38\textwidth}
  \includegraphics[width=\linewidth]{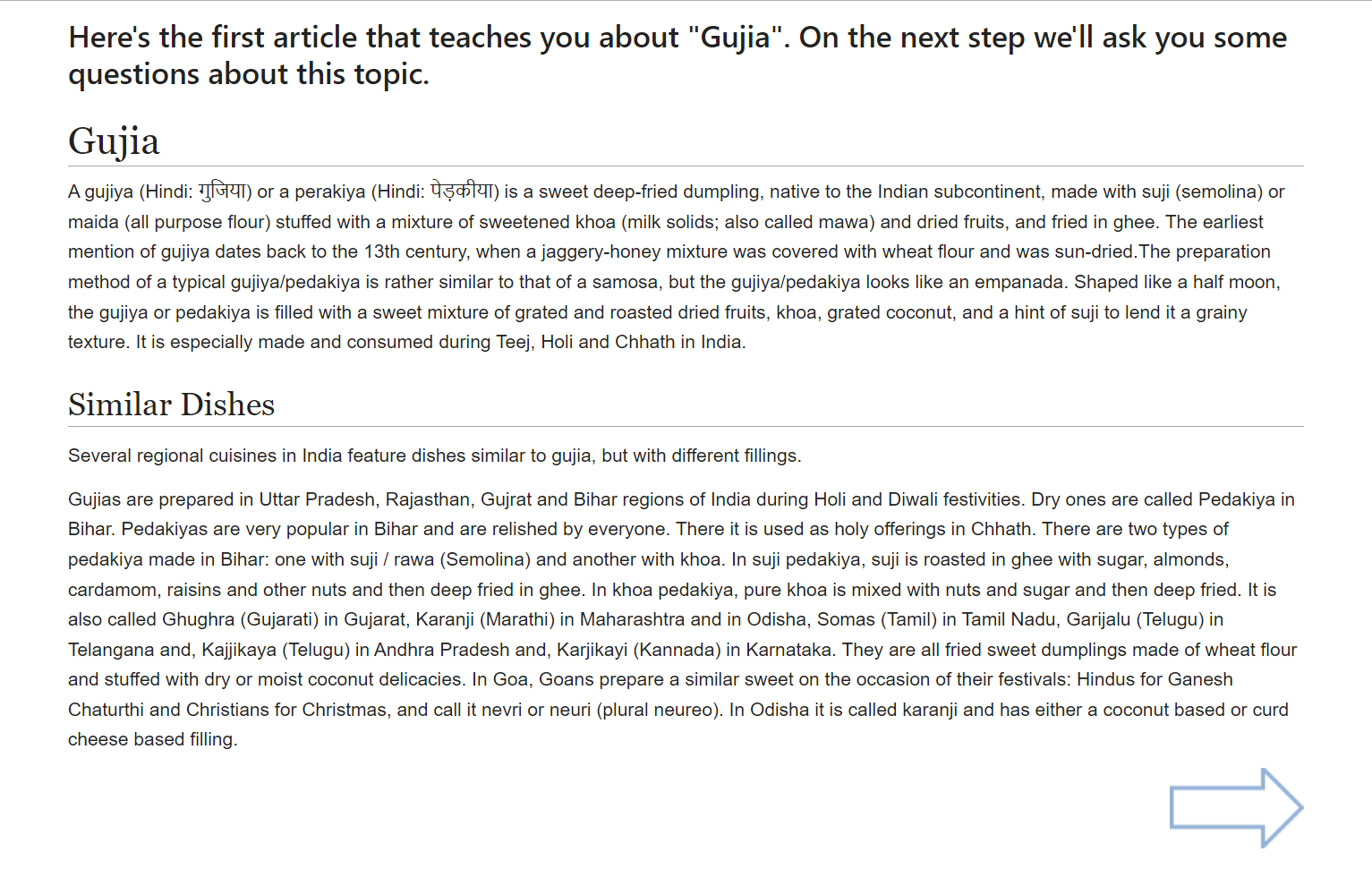}
  \caption{Article displayed without an image}\label{fig:article_w_out_img}
\end{subfigure}
\caption{Examples of articles with/without images shown in our experimental setting.}
\end{figure*}
Because articles could vary significantly in detail and length, we created excerpted versions. The excerpt included the lead (top) section of the article and any section that was used to generate general or visual knowledge questions. For example, the answer to ``The P\`ere David's deer is significant to which cultural mythology?'' was found under the ``Legend and cultural significance'' subheading. This subsection was thus incorporated into the article's excerpt (for this specific article, the except also included the Characteristics subsection). At most an except would be comprised of 4 subsections (one for question). However, most of the excerpts were made up of two sections (median, mean=2.38) with a mean/median word count of 373 and 359, respectively. Using the Flesch-Kincaid grade level metric, we found the median readability grade level for our excerpts was 12 (M = 11.84, SD = 1.98, min = 7, max = 17). This is in line with the readability distributions observed in the past for Wikipedia~\cite{lucassen2012readability}.
 
\paragraph{Final Dataset}

In summary, a total of 94 articles were used to create the dataset. The data includes the excerpted article and the `lead' image (the header/top image seen by visitors to Wikipedia to that article). A total of 470 questions were produced, with five questions per article (two general knowledge questions, two visual knowledge questions, and one image recognition question). We provide this dataset in our supplemental materials (\url{https://github.com/datamazelab/wikiimages}).
\section{Online Experiments}

Our online experiment was primarily aimed at answering: [RQ1] Do Wikipedia articles with images help participants answer questions more accurately compared to articles without an image? and [RQ2] What characteristics of the images are most helpful in answering the different question types?

Our baseline experiment measures participant performance without the presence of Wikipedia article excerpts and images. For the main experiment, we designed a within-subjects online experiment using our full knowledge comprehension dataset.
 The experiments were approved by the Institutional Review Board at the University of Washington. The main experiment was preregistered on AsPredicted (\url{https://aspredicted.org/JJQ_GVQ}).  We slightly rephrased the preregistered research question for conciseness and added RQ2, which the preregistration did not explicitly list (though it was implied through the types of analyses).

\subsection{Baseline Experiment}\label{sec:baseline}

Our baseline experiment was designed to determine the difficulty of the questions in our dataset without showing the content or images of the article.

The baseline experiment began with an informed consent form and a brief overview of the study, before participants were asked 12 questions total. Six (6) questions belonged to one article with one (1) familiarity question (prior experience with the article's topic), two (2) general knowledge questions, two (2) visual knowledge questions, and a final image recognition question (1). The familiarity (article knowledge) question included five options ranging from ``I have never heard of it'' to ``I am an expert on this topic.'' A second set of six questions followed the same format, but were asked on a different article topic. Questions were asked without participants seeing the Wikipedia article.  Participant responses were collected for all questions associated with the 94 articles in our dataset. All collected metrics are described in Section \ref{sec:metrics}. 

\begin{figure*}[!tb]
\minipage{0.9\textwidth}
  \includegraphics[width=\linewidth]{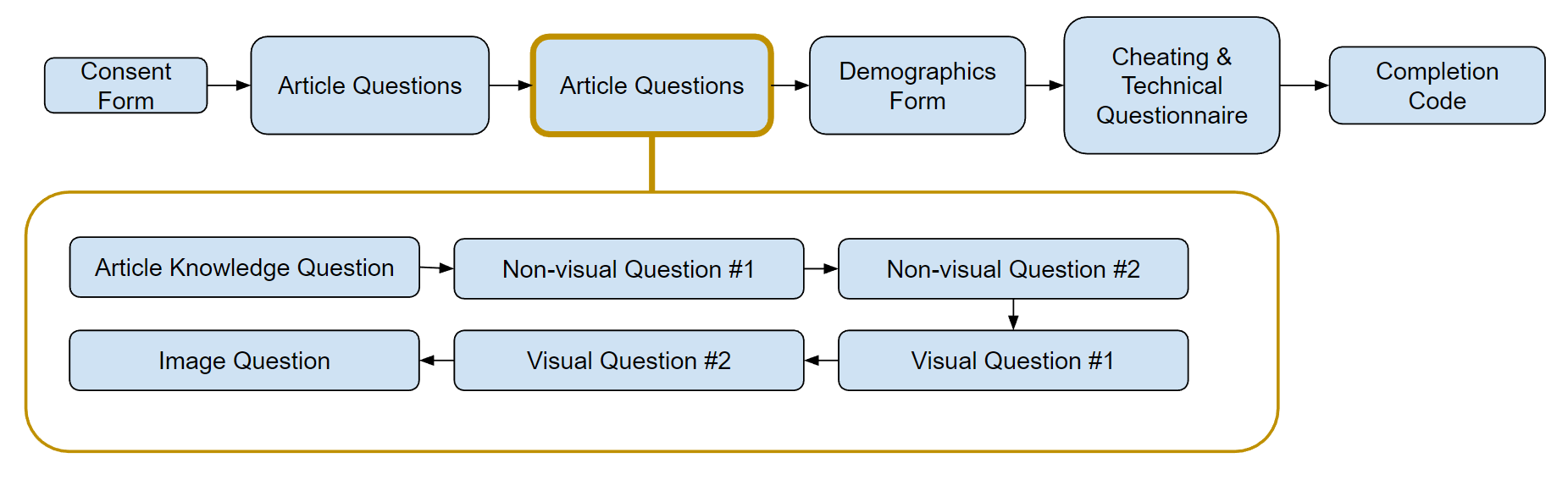}
  \caption{Procedure Flow Diagram for the Baseline  Study}\label{fig:procedure_two}
\endminipage\hfill
\end{figure*}

\begin{figure*}[!tb]
\minipage{0.9\textwidth}
  \includegraphics[width=\linewidth]{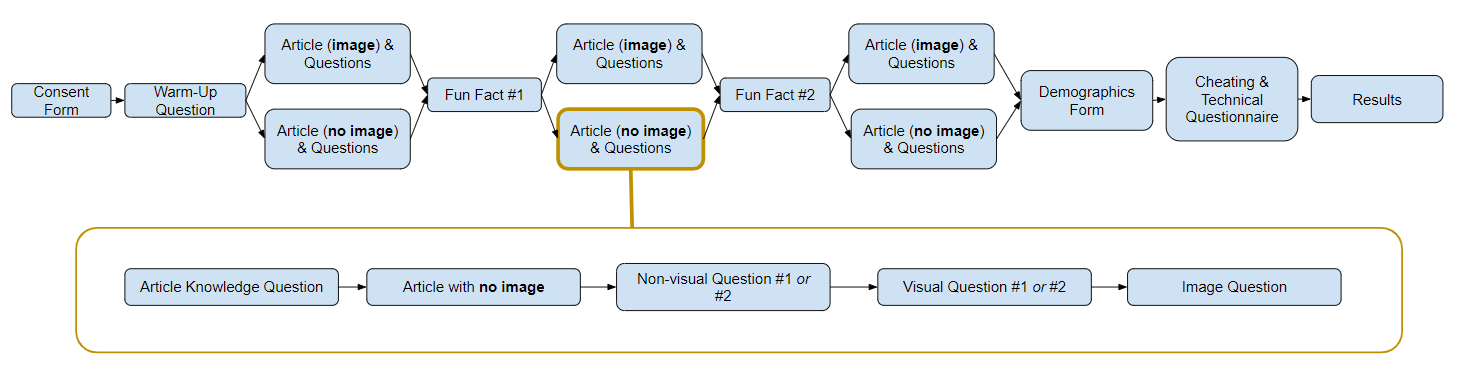}
  \caption{Procedure Flow Diagram for our main experiment}\label{fig:procedure_one}
\endminipage\hfill
\end{figure*}

\subsection{Participants}
\paragraph{Participants} We recruited participants from Prolific, an online research study platform. A total of 51 Prolific participants completed the baseline experiment.  Each received \$0.25/min (\$15/hr) for their participation. Only 47 data entries out of 51 were used (four participants indicated that their data should not be included, e.g., because of self-acknowledged cheating). Of the 47, 47.82\% self-identified as female, 50\% as male, and 2.17\% participants answered using other expressions (e.g., non-binary, bigender, etc.). The max age of the participants was 69, and the minimum age recorded was 18 (M = 36.23, SD = 12.35). Two different countries of origin were reported by the participants: United Kingdom (91.5\%), and United States (8.5\%). A majority  (91.11\%) were English speakers, 4.44\% Lithuanian, and 4.44\% German speakers. The distribution of English proficiency was 86.95\% advanced, 10.86\% upper-intermediate, and 2.17\% intermediate. The distribution of education levels was 21.73\% high school, 26.08\% college, 10.86\% professional school, 34.78\% graduate school, and 6.52\% Ph.D. Participants took at most 7 minutes in the study and at least 1.5 minutes (M = 3.11, SD = 1.06).

\subsection{Main Experiment}\label{sec:main_exp}
The main experiment began with an informed consent form and a brief overview of the study, as seen in the upper part of Figure~\ref{fig:procedure_one}. A practice trial asked participants to try their best and answer the randomly chosen general knowledge, visual knowledge, and image recognition questions seen on the same page. This allowed all participants to see the range of questions they might be asked during the experiment. As with the baseline experiment, participants were asked about their familiarity with a randomly chosen Wikipedia article topic on the following page. 

The main part of the experiment consisted of three trials, each including two parts: reading excerpts from a Wikipedia article and answering three questions for each article on the following pages.  The articles were randomly selected from our dataset and randomly shown with or without images. Before displaying each article, the familiarity question was asked.

Participants received instructions to read the article and that they would be asked questions about it on the next page, without having access to the article at that point. This prevented participants from being able to refer to the content while being assessed on their retention. The pages used to display the excerpted article mimicked the design of standard Wikipedia articles (e.g., Figure \ref{fig:article_w_img} and Figure \ref{fig:article_w_out_img}). The study display was accessible to different device users through a feature that responded to different screen sizes.

There was a 50\% chance for a participant to see an image with their article. If a Wikipedia article was shown with the image, we used the image from the original Wikipedia article in its original resolution, preloading it to avoid viewing delays that would impact our time measures.  Images were displayed in a location similar to their placement on the original Wikipedia pages (on the top right of the article for tablet and computer users or centered on the top for smaller device users, see Figure \ref{fig:article_w_img}). 

The three comprehension questions (general knowledge, visual knowledge, and image recognition) were displayed on three separate pages. All three were multiple choice and included up to four answer options. Recall that when generating the question set, we created two general and two visual knowledge questions per article as discussed in Section \ref{sec:q_generation}. For these question types, we randomly chose to display one of the two choices on their respective pages.

Participants received instant visual and audio feedback on their general knowledge, visual knowledge, and image recognition question selections. 
 In between trials, participants were given the option to take a break while reading a random fun fact taken from the Wikipedia `unusual articles page'~\cite{unusualWiki}. In total, each participant completed twelve questions, four for each of the three articles (familiarity/article knowledge question, general knowledge question, visual knowledge question, and an image recognition question).

The collected data from this experiment contains unevenly distributed answers per question. Each participant was randomly presented with three articles (out of 94 total articles) and three questions per article (out of 5 possible). After each article that participants viewed, a page with a general knowledge question and a page with a visual knowledge question were randomly displayed. The general and visual knowledge questions were randomly selected from two generated options for each type. Due to the random selection of questions, even if two participants responded to questions created for the same article, they did not necessarily answer the same visual and general knowledge questions for that article. 

Collected data for the main experiment includes question responses to 90 articles out of 94. A total of 368 question responses were collected for 470 generated questions, excluding familiarity (article knowledge) questions. The average number of answers collected per question is 18.29 (SD$=$11.23, max$=$74 and min$=$2).

At the end of the experiment, we asked participants to answer demographic questions about their level of education, country of origin, age, gender, native language, if they have done the study before, and English reading proficiency. The next page asked if they had technical issues or cheated at any point in the study to determine whether their data should be excluded. Finally, they saw a results page including a personalized ``Wiki Knowledge Score'' that took into account how long they took and how well they answered each question compared to others. 
An extra slide was used for Prolific participants to get their Prolific ID and to display a completion code.
The study took up to 10 minutes to complete and was entirely in English.

The experiment design deviated slightly from our preregistered design. Upon initial analysis, we found removing the data collected for the first article's questions necessary to account for participants getting used to the experiment structure. We further discuss excluding this data in the Results section and provide the results when the first article's questions are included in the supplemental files.

We launched the experiment on LabintheWild~\cite{labinthewild}, a volunteer-based experiment platform where participants receive personalized feedback in exchange for study participation, and on Prolific, where participants receive financial compensation. This dual-platform strategy had two functions: (1) it allowed us to quickly recruit a sufficient number of participants to support our statistical analysis, and (2) the two populations roughly mimic the mix of intrinsic and extrinsic motivators of those reading Wikipedia (e.g., personal curiosity and grades, or in the case of Prolific, money). Since LabintheWild participants are intrinsically motivated to test themselves and perform well, LabintheWild studies have been repeatedly shown to result in higher data quality than studies on paid online recruitment platforms, with participants paying more attention, providing more consistent responses and scoring higher~\cite{ye2017personalized,august2019pay}. Prolific participants were from the US and UK and were paid 50 cents per article/question set (yielding roughly \$15/hour).

A total of 748 participants attempted the experiment, 445 through LabintheWild, and 303 through Prolific. 
We excluded 37 LabintheWild participants from the data set. These were removed if they completed the study in less than a minute (13), or indicated reasons for excluding the data (e.g., self-reported cheating) (8) or technical issues (16). The latter two variables are routinely collected in LabintheWild experiments to give participants an option to indicate that their data should be excluded from analysis. Collecting the same information on Prolific, we also removed 7 out of 303 Prolific participants (six for technical issues and one for finishing in under a minute). This resulted in a total of 704 participants,  408 (57.95\%) from LabintheWild and 296 (42.04\%) from Prolific.

\begin{figure}[t]
\centering
\begin{subfigure}[b]{0.4\textwidth}
  \includegraphics[width=\linewidth]{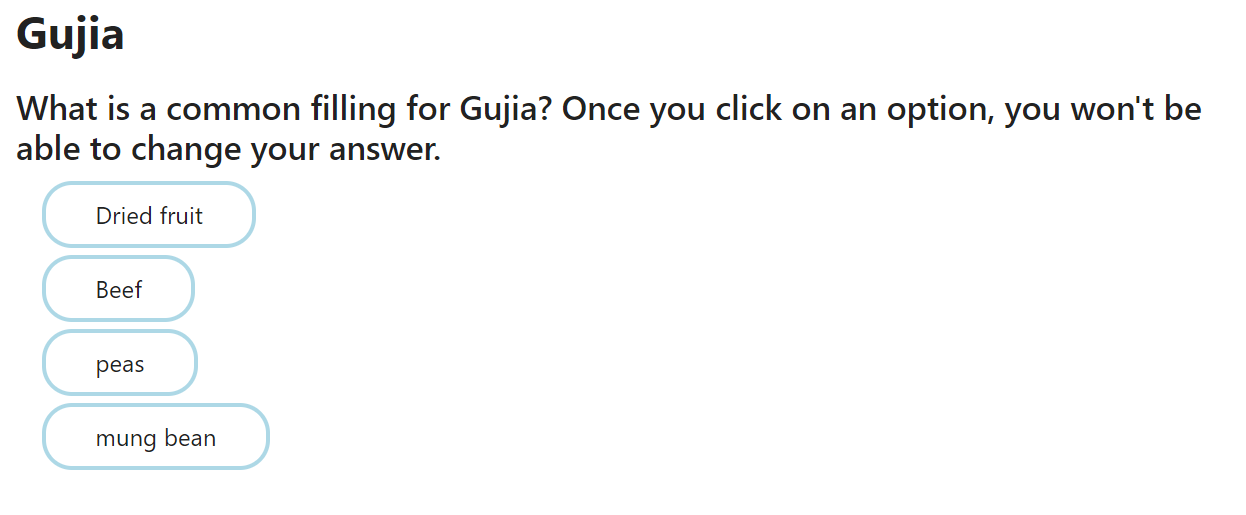}
  \caption{General knowledge question}\label{fig:non_vis}
\end{subfigure}
\begin{subfigure}[b]{0.4\textwidth}
  \includegraphics[width=\linewidth]{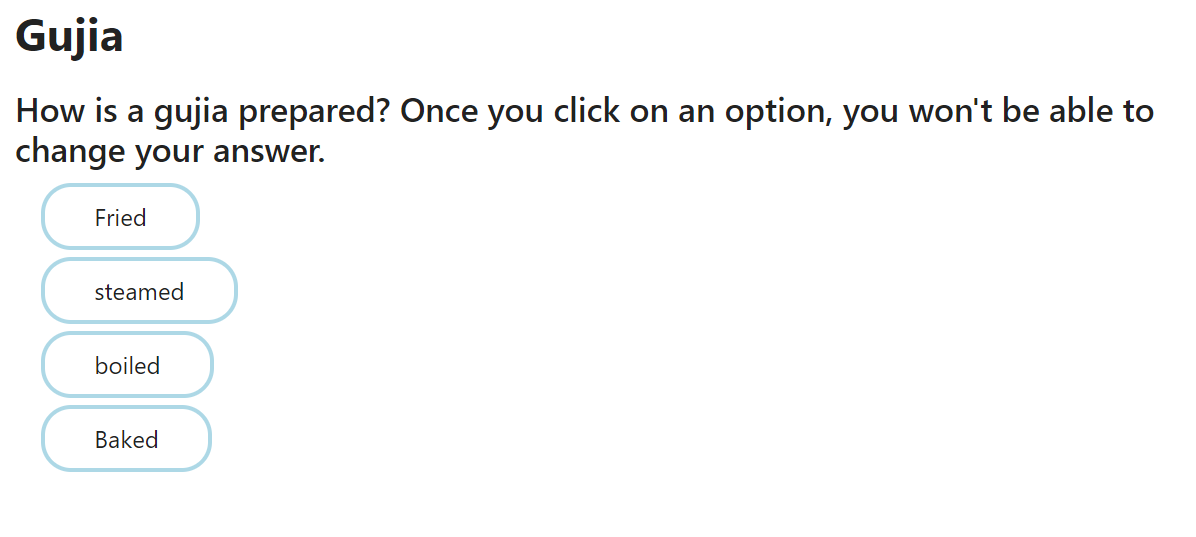}
  \caption{Visual knowledge question}\label{fig:vis}
\end{subfigure}
\begin{subfigure}[b]{0.4\textwidth}
  \includegraphics[width=\linewidth]{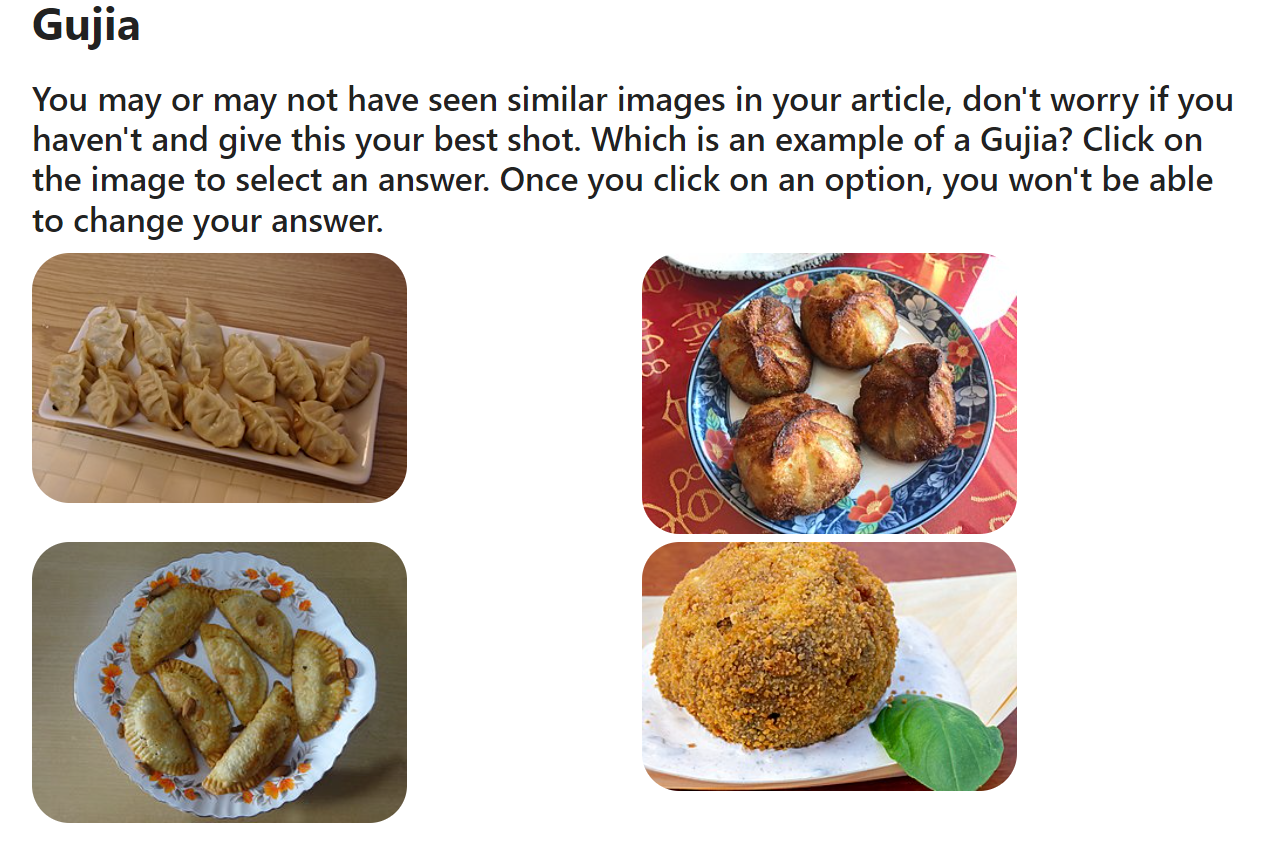}
  \caption{Image recognition question}\label{fig:img}
\end{subfigure}
\caption{Examples of (a) General knowledge questions, (b) Visual knowledge questions and (c) Image recognition questions from our experimental setting.}
\end{figure}
Among our 704 participants, 47.57\% self-identified as female, 48.71\% as male, and 3.70\% participants answered using other expressions (eg. non-binary, bigender, etc.).  Participants were between 10 and 89 years old (M = 32.23, SD = 23.60). In total, 53 different countries of origin were reported by the participants, the top 5 being the United Kingdom (44.79\%), United States (29.10\%), Canada (4.27\%), Germany (3.28\%), Australia (2.13\%), and France (1.42\%). Participants indicated being native speakers of 31 different languages, the majority being English (77.23\%). The distribution of English proficiency was 82.91\% advanced, 12.26\% upper-intermediate, 3.06\% intermediate, 1.02\% beginner, and .07\% basic. The distribution of education levels was 3.43\% pre-high school, 15.04\% high school, 35.81\% college, 9.31\% professional school, 27.07\% graduate school, 7.16\% PhD, and 2.14\% postdoctoral. Participants took at most 28 minutes on the study and at least 2 minutes (M = 8.59, SD = 4.22). 

\subsection{Metrics}\label{sec:metrics}
For each participant, article, and question, we recorded several metrics for our analysis.

\subsubsection{Participants' Characteristics}
We stored for each participant:
\begin{itemize}
\item \textbf{Participant ID}, recorded anonymously using UUID format.
\item \textbf{English Proficiency}, taken from the demographics form, can take one of the following values: \textit{beginner}, \textit{basic}, \textit{intermediate}, \textit{upper-intermediate}, and \textit{advanced}.
\item \textbf{Education Level}, taken from the demographics form, whose values can be: \textit{pre-high school}, \textit{high school}, \textit{college}, \textit{graduate school}, \textit{professional school}, \textit{postdoctoral}, \textit{PhD}.
\item \textbf{Age} was taken from the demographics form.
\item \textbf{Platform}, either \textit{LabintheWild}, or \textit{Prolific}. LabintheWild was used for participants that organically found and took the study through LabintheWild platform. While Prolific was used for participants that took the experiment for paid compensation using the Prolific platform.
\end{itemize}

Additionally, metrics for gender, country of origin, reasons for data exclusion (e.g., self-reported cheating), and indication of technical issues were recorded from the demographics form and the self-reported questionnaire at the end of the experiment. Devices and browsers used by the participants were also logged after participants agreed to the consent form. 

\subsubsection{Article Properties}
Article metrics included:
\begin{itemize}
    \item 
\textbf{Article Id}, a unique ID--one for each of the 94 articles
\item
\textbf{Article Time} corresponds to the time taken by a participant to read or view an article\footnote{before they clicked a blue arrow to see the next slide which included the first question for that article}.
\item
\textbf{Article Knowledge}, the answer to a five-nominal scale familiarity question regarding the topic of the article. The scale categories were: ``I’ve never heard of it'', `I’ve heard of it but I don’t know what it is'', `I have some knowledge about this topic'', `I know a lot about this topic'', and `I’m an expert on this topic''.
\item
\textbf{Article Word Count} for each of the 94 articles in our dataset.
\item
\textbf{Image Presence} a binary variable corresponding to the presence/absence of an image in an article viewed by a participant.
\item
\textbf{Readability} was determined using the Flesch-Kincaid grade-level metric. \end{itemize}

\subsubsection{Question Properties}
We record the following data for questions in our experiment:
\begin{itemize}
    \item 
\textbf{Question Type} identifies the type of question the participant saw: \textit{general}, \textit{visual}, or \textit{image}. 
\item
\textbf{Individual Question Score}, a binary variable that identifies whether the question was answered correctly or not.
\item
\textbf{Question Time}, the difference between timestamps of when the participant first saw the question and when they clicked a blue arrow to see the next slide in the experiment.
\item
\textbf{Visual Knowledge Question Consistency}, an indicator (only for visual knowledge questions) of whether the question could be answered by looking at the image alone.
\end{itemize}

\subsection{Models}\label{sec:models}
To understand how images impact participant question accuracy, we fitted a series of linear mixed-effects models using the data collected from the main experiment and the lme4 R package \cite{lme4}. To train these models, we excluded data from the 101 participants who chose not to provide optional demographic information such as age, gender, education level, and English proficiency. This resulted in the models using data from 603 participants. We separated our data into three sets (i.e., by  ``general'', ``visual'', and ``image'' recognition question responses) and trained three separate full and null models using the separated question type data sets. 

First, the null models (intercept/empty) were fitted with \textit{Article Id} and \textit{Participant Id} as random factors and \textit{Individual Question Score} as the dependent variable. Next, we created full models with \textit{Individual Question Score} as the dependent variable, and we used  the following as independent variables: \textit{Article Time}, \textit{Question Time}, \textit{Article Knowledge}, \textit{Age}, \textit{Article Word Count}, \textit{English Proficiency}, \textit{Education Level}, \textit{Platform}, and \textit{Image Presence}. Again, \textit{Article Id} and \textit{Participant Id} were used as random factors. The visual question model includes an extra independent variable of \textit{Visual Knowledge Question}. We used the null and full model created for each data set category (``general'', ``visual'', and ``image'') to perform comparison using the
likelihood ratio test.

Each full model produced unique results showing how distinct parts of the experiment impacted participants' success in answering the three different question types. These models were built on limited data and our main focus was not on achieving better model prediction. Instead, we wanted to identify variables within our experiment that influenced participants' knowledge acquisition which we gauged through question accuracy. Section \ref{sec:results} further explores our results, showing that each full model for the different question types contains a marginal $R^2$ below $.05$. Clearly, there may be other uncaptured variables that better predict question accuracy. Nonetheless, these models give relevant insights into the variables that affected participants in our experiment.

\section{Results}\label{sec:results}

Our results show that images in Wikipedia articles are only helpful in some cases of knowledge acquisition. The presence of an image did not significantly affect the accuracy of the responses if participants were asked general and visual knowledge questions. However, it significantly improved the accuracy of the responses to the image recognition questions. 

\subsection{Baseline and (No) Image Performance}\label{sec:baselineres}
\begin{figure}
    \centering
    \includegraphics[width=\linewidth]{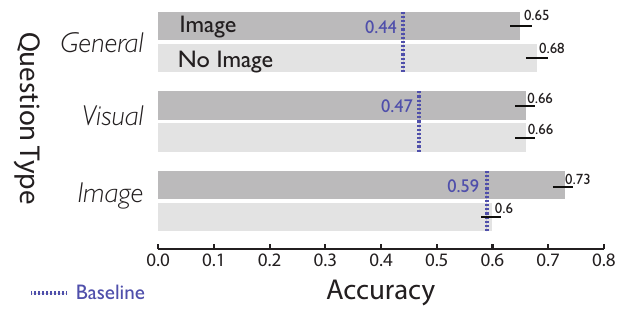}
    \caption{All question types broken into the image + text condition (``Image'') and the text-only conditions (''No Image''). These are compared to baseline performance (blue line) when neither text nor image are provided.}
    \label{fig:all_qs_baseline_comparison}
\end{figure}

To establish whether learning could take place, and how much, we began by analyzing the data from our baseline experiment (see~\ref{sec:baseline}). The average test score for each type of question, without exposure to article text nor images, indicates both how well an `average' participant knows the material and how guessable/inferable the correct answer is. Baseline accuracy was 44\%, 47\%, and 59\% for the general knowledge, visual knowledge, and image recognition questions respectively (see Figure~\ref{fig:all_qs_baseline_comparison}). This is higher than a random guess (25\%) but does not create a ceiling effect.

As can be seen in Figure~\ref{fig:all_qs_baseline_comparison}, any intervention (text or text with image), results in significant improvements over baseline for knowledge acquisition and retention ($F_{(1)}=60.91, p < .0001$). Including an image with the article helps with the image recognition questions (73\% versus 59\%). Interestingly, with text alone, participants did not perform much better than the baseline for these questions (60\% versus 59\%). This supports the critical importance of images for some learning tasks.

On initial analysis, images did not appear to help with general knowledge questions and may hurt performance slightly (though the difference is not significant). With general knowledge questions, there is no `knowledge' benefit to the images, so this result is perhaps expected. However, there is always a concern that images may be a distraction and lead to less engagement with the text. More surprisingly, there were no differences in the visual knowledge questions with and without the image (66\%). We expand on these results below.

\subsection{Do images help learning? (RQ1)}\label{sec:rq1}
To test the impact of images in question accuracy, we applied regression models using various metrics collected for individuals, questions, and articles. Below we describe each of the question type models in more detail.

\subsubsection{General Type Question Model}\label{sec:non-vis_q}
\begin{figure}[!htb]
\begin{subfigure}{0.48\textwidth}
  \includegraphics[width=\linewidth]{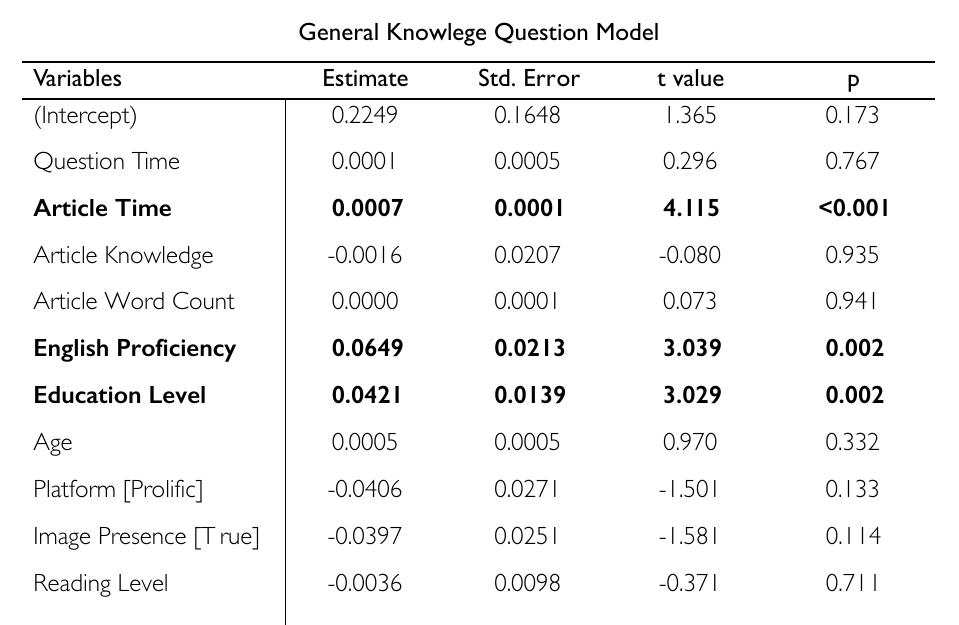} 
  \caption{Mixed effect model predicting general knowledge question type accuracy. Adjusted $R^2 = .19$. }\label{fig:non_vis_table}
\end{subfigure}
\hfill
\begin{subfigure}{0.48\textwidth}
    \includegraphics[width=\linewidth]{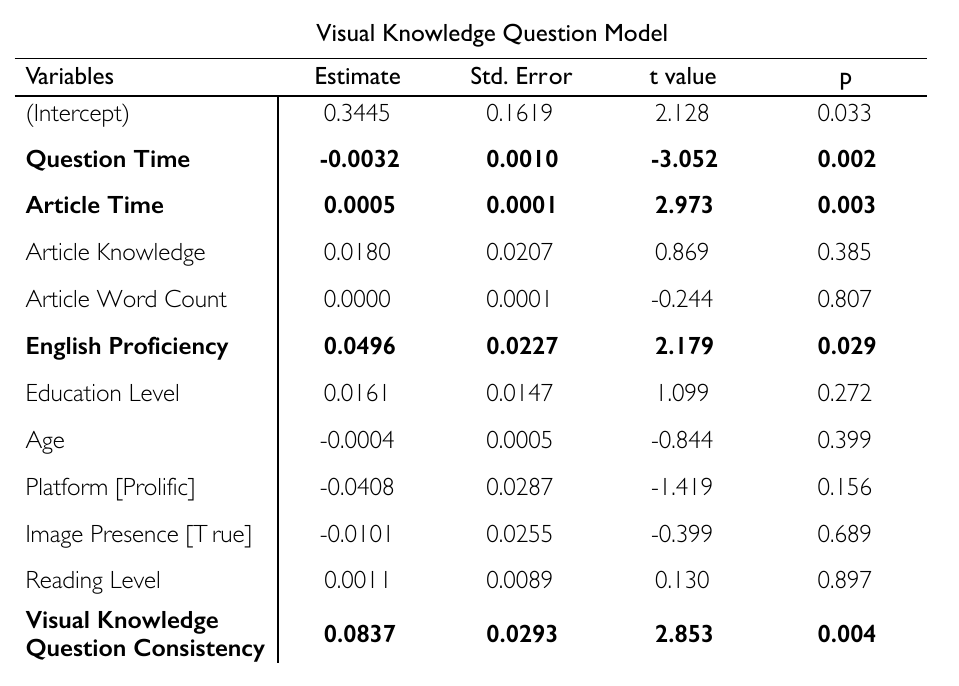}
  \caption{Mixed effect model predicting visual knowledge question type accuracy. Adjusted $R^2 = .209$. }\label{fig:vis_table}
\end{subfigure}
\begin{subfigure}{0.48\textwidth}
  \includegraphics[width=\linewidth]{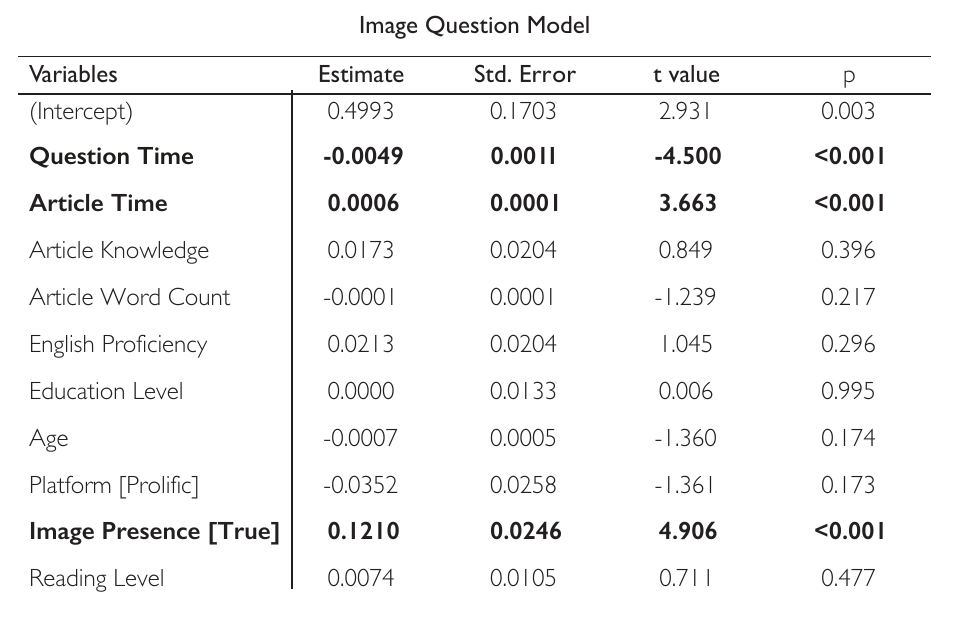}
  \caption{Mixed effect model predicting image recognition question type accuracy. Adjusted $R^2 = .221$.}\label{fig:image_table} 
\end{subfigure}
\caption{Results for the (a) General Knowledge, (b) Visual Knowledge, and (c) Image Recognition question type models}
\end{figure}

A comparison of the null model with the full general question type model using the
likelihood ratio test showed that the full model fits the data significantly better ($\chi^2_{(15)} = 2593.7$, $p < .0001$). Our General Knowledge Question Model only looked at the question data collected from general knowledge questions and explains 19\% of participants' performance on these specific questions (conditional $R^2=.19$, marginal $R^2=.034$). As can be seen in Fig.~\ref{fig:non_vis_table}, the presence of the image does not significantly improve the question score for the general knowledge question type.  Specifically, the model indicates that the longer participants read or viewed the article, the better they performed on the general knowledge question ($F_{(1)}=16.93, p < .001$). Participants also did better on general knowledge questions when they indicated that they had a higher level of education ($F_{(1)}=9.24, p < .05$) and English proficiency ($F_{(1)}=9.17, p < .05$). Readability level and image presence did not have a statistically significant effect on knowledge comprehension for general knowledge questions.

\subsubsection{Visual Knowledge Question Type Model}\label{sec:vis_q}
A comparison of the null model with the full visual knowledge question type model using the
likelihood ratio test showed that the full model fits the data significantly better ($\chi^2_{(15)} = 2678.0$, $p < .0001$). We created the Visual Knowledge Question model using data collected for the visual knowledge question type. The model explains 20.9\% of participants' performance (conditional $R^2=.209$, marginal $R^2=.025$). The model shows that question time, article time, English proficiency, and visual knowledge question consistency affect how well people did on visual knowledge questions. The longer the time that participants took to answer the visual knowledge question, the lower their question score  ($F_{(1)}=9.31, p < .05$) as seen in Fig. \ref{fig:vis_table}. Similarly to the General Knowledge Question Type Model, the longer participants took viewing the given article the better they did on the visual knowledge question as well ($F_{(1)}=8.84, p < .05$). This model also indicated that English proficiency had a significant main effect on participant accuracy ($F_{(1)}=4.3, p < .05$). Similarly to general knowledge questions, readability level and image presence had no statistically significant effect on visual knowledge questions. For visual knowledge questions, we included an additional variable of visual consistency. In the resulting model, the variable indicates that visual 
 knowledge questions that \textit{could} be answered with the help of the randomly shown image (i.e., were consistent) yielded better performance ($F_{(1)}=8.14, p < .05$).

\subsubsection{Image Recognition Question Type Model}\label{sec:image_q}
A comparison of the null model with the full image recognition question type model using the
likelihood ratio test showed that the full model fits the data significantly better ($\chi^2_{(15)} = 2513.5$, $p < .0001$). Image Recognition Question Type Model that only looked at data for the image recognition question type resulted in explaining 22.1\% of participant's question score on the image recognition question (conditional $R^2=.221$, marginal $R^2=.048$). Both lower question time ($F_{(1)}=20.24, p < .001$) and longer article time ($F_{(1)}=13.41, p < .001$) increased participant knowledge comprehension as seen in Fig. \ref{fig:image_table}. Out of all the models, this is the only one where the presence of the image significantly impacted the accuracy of the participants. With images present, participants performed significantly better on the image recognition question ($F_{(1)}=24.06, p < .001$).

Recall that for all experiments we treated the first article as a practice round and excluded it from analysis (Section \ref{sec:main_exp}).  Notably,  after removing the first article data, our model remained largely unchanged for all question types with one key exception: the platform became significant. Individuals on the Prolific platform appeared to do worse overall. We hypothesize that this may have to do with motivation. LabintheWild participants, who were more intrinsically motivated, may be engaged from the first article and question. Prolific participants may optimize largely for specific task success. Once they have learned the task, they will focus on reading to achieve better results. Future work may validate this hypothesis or provide alternative explanations.

\subsection{Which images help learning? (RQ2)}\label{sec:rq2}
\begin{figure*}
    \centering
    \captionsetup{justification=centering,margin=4mm}
    \begin{subfigure}[t]{0.6\textwidth}
    \begin{subfigure}[t]{0.32\textwidth}
        \includegraphics[width=\textwidth]{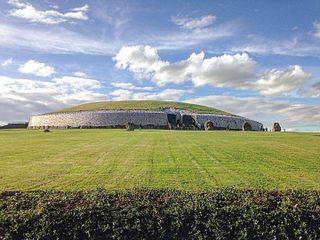}
        \caption*{\footnotesize \\During which period was Newgrange built?\\}
    \end{subfigure}
    \begin{subfigure}[t]{0.32\textwidth}
        \includegraphics[width=\textwidth]{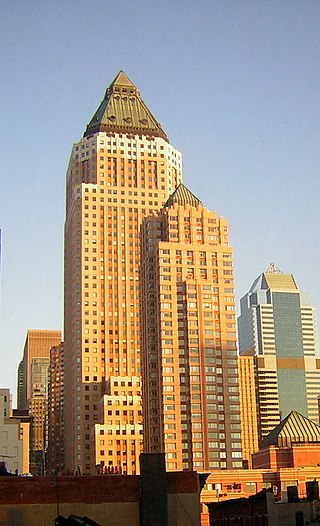}
        \caption*{\footnotesize \\When was the Worldwide Plaza completed?\\}
    \end{subfigure}
    \begin{subfigure}[t]{0.32\textwidth}
        \includegraphics[width=\textwidth]{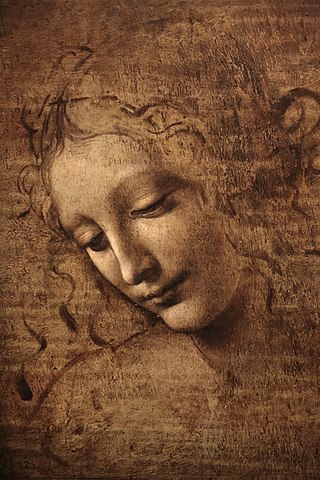}
        \caption*{\footnotesize \\When was "La Scapigliata" painted?}
    \end{subfigure}
    \caption{General knowledge questions where images help with content retention.}
    \end{subfigure}
    ~ 
    \begin{subfigure}[t]{0.4\textwidth}
    \begin{subfigure}[t]{0.47\textwidth}
        \includegraphics[width=\textwidth]{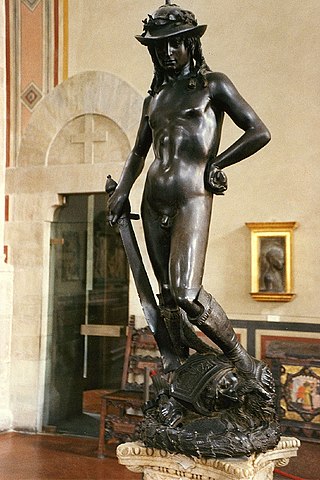}
        \caption*{\footnotesize \\What is the material of Donatello’s David?}
    \end{subfigure}
    \begin{subfigure}[t]{0.47\textwidth}
        \includegraphics[width=\textwidth]{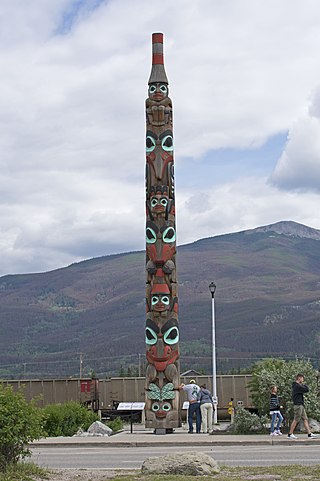}
        \caption*{\footnotesize \\How tall can totem poles be?}
    \end{subfigure}
    \caption{Visual knowledge questions where images help content retention.}
    \end{subfigure}
    \captionsetup{justification=justified}
    \caption{Qualitative analysis--examples of images that are helpful to answer generic questions. (a) Visual arts characteristics for general knowledge questions. (b) Visually consistent images for visual knowledge questions.}
    \label{fig:nonimagequesiton}
\end{figure*}

The literature studying the role of images in instructional settings suggests that the impact of images on knowledge acquisition depends, among other things, on the visual characteristics of the image (extensively surveyed by Clark et al.~\cite{clark2010graphics}).

To understand whether the trends outlined above were consistent across different image types, we performed a manual inspection of the images whose presence generated greater positive/negative differences in the question scores.

\subsubsection{General knowledge questions}
Our analysis shows that images largely had no significant impact on general knowledge question accuracy. To expose interesting patterns of exception to this trend, we inspected our results more in-depth. For each question, we have a \textit{question score} that is either 0 if the question was answered correctly (1, otherwise).  This makes scores quite comparable across questions of different types. For each question, an \textit{image effect} value was calculated, by averaging the \textit{question score} for all responses where \textit{image presence} was true, and subtracting the average \textit{question score} for all responses where \textit{image presence} was false. The larger the image effect value, the larger the impact of the image presence on the participant's accuracy.  Image effect scores ranged from $-1$ for questions where image presence had negative impact, all the way to $0.83$ for questions where images have a positive impact on accuracy. We found that, as expected, the image effect score is 0 for the vast majority of questions (65\%), and that only a small number of questions have a non-zero image effect, equally distributed across the positive and negative spectrum. We then ranked general knowledge questions by image effect, and examined the top 15 questions, articles, and images, for which the image had a positive effect. We found that, in some cases, images were useful to answer general knowledge questions for articles about visual arts and architecture. Images were helpful to estimate epochs (i.e., time periods) and styles of paintings, or the century when monuments were built (see Fig. \ref{fig:nonimagequesiton}(a)).

\begin{figure}[t!]
    \centering
    \includegraphics[width=\linewidth]{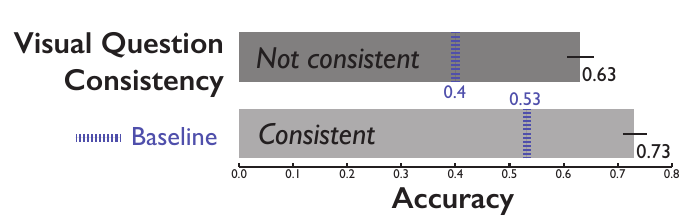}
    \caption{Visual Knowledge Question Compared to Baseline Performance with Consistency Metric}
    \label{fig:vis_q_baseline_comparison}
\end{figure}

\subsubsection{Visual knowledge questions}
As indicated by our high-level analysis, we saw no effect of images on visual knowledge questions. Although we used only text to create visual knowledge questions, when present, images should have been helpful in answering these questions, especially those images that are consistent with the visual descriptions in the text.  To better understand the impact of images on visual knowledge question accuracy we labeled each image in our dataset with a ``visual knowledge question consistency'', that indicated whether a visual knowledge question could be answered using the article image (see Figure \ref{fig:nonimagequesiton}(b)). In Figure \ref{fig:vis_q_baseline_comparison} we can see how visual knowledge question consistency impacted accuracy in comparison to the baseline. The baseline accuracy between consistent and not consistent visual knowledge questions is not significant ($F_{(1)}=2.8, p = .096$) based on existing limited data.  However, consistency is a significant main effect ($F_{(1)}=6.32, p < .05$) for the visual knowledge question type data collected from our main experiment. Participants tended to get correct answers to visual knowledge questions more often when consistent images are displayed along with the article. 

\begin{figure*}
\centering
\captionsetup{justification=centering}
    \begin{subfigure}[b]{0.32\textwidth}
    \begin{subfigure}[b]{0.48\textwidth}
        \includegraphics[width=\textwidth]{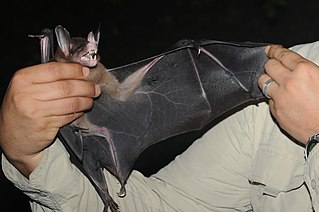}
        \caption*{\footnotesize Low quality  (\textit{Spectral Bat})}
    \end{subfigure}
    \begin{subfigure}[b]{0.48\textwidth}
        \includegraphics[width=\textwidth]{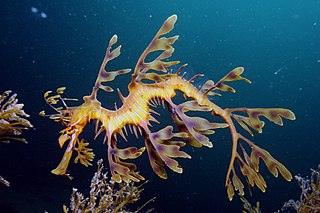}
        \caption*{\footnotesize High quality  (\textit{Leafy Seadragon})}
    \end{subfigure}
    \caption{Image quality}
    \end{subfigure}\hfill
    ~ 
    \begin{subfigure}[b]{0.32\textwidth}
    \begin{subfigure}[b]{0.48\textwidth}
        \includegraphics[width=\textwidth]{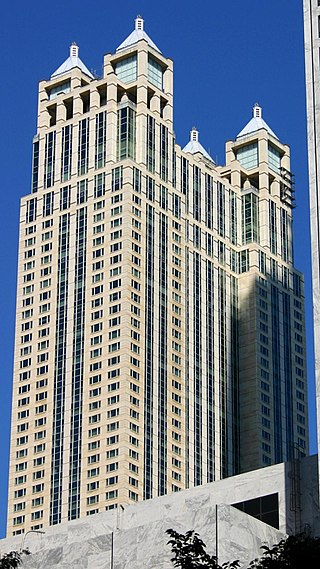}
        \caption*{\footnotesize Part (\textit{900 North Michigan})}
    \end{subfigure}
    \begin{subfigure}[b]{0.48\textwidth}
        \includegraphics[width=\textwidth]{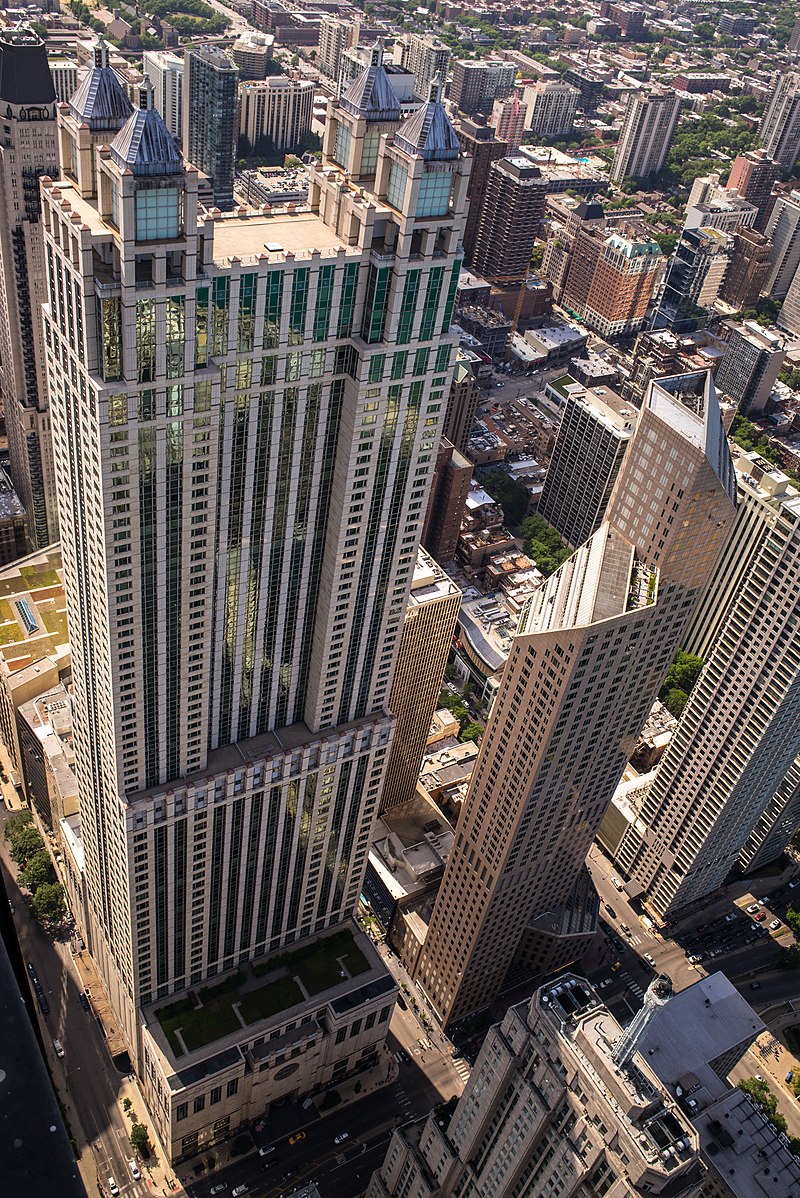}
        \caption*{\footnotesize Whole (\textit{900 North Michigan})}
    \end{subfigure}
    \caption{Part VS Whole}
    \end{subfigure}\hfill
    \begin{subfigure}[b]{0.32\textwidth}
    \begin{subfigure}[b]{0.48\textwidth}
        \includegraphics[width=\textwidth]{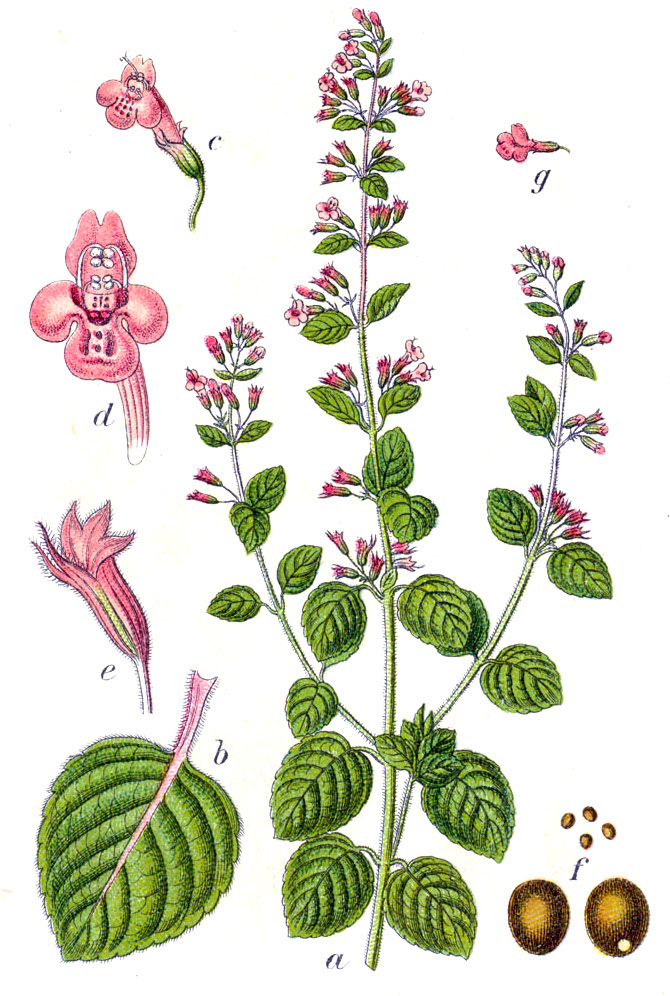}
        \caption*{\footnotesize Illustrated\\ version (\textit{Nepeta})}
    \end{subfigure}
    \begin{subfigure}[b]{0.48\textwidth}
        \includegraphics[width=\textwidth]{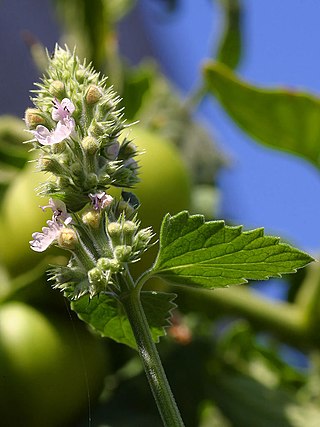}
        \caption*{\footnotesize Photographed version (\textit{Nepeta})}
    \end{subfigure}
    \caption{Photo VS Illustration}
    \end{subfigure}
    \captionsetup{justification=justified}
    \caption{In-depth analysis - examples of images that are harmful to answer the image question. (a) Low Quality images. (b) Images that describe only a part of the object. (c) Photographed objects in the article vs illustrations in the question.}
    \label{fig:imagequestion}
\end{figure*}

\subsubsection{Image recognition questions}
In our quantitative results, we observed that participants who saw text+image articles generally performed significantly better on image recognition questions. To reveal possible exceptions to this trend, we performed the same ranking procedure of general knowledge questions and inspected the 15 questions ranked low by image effect. The image effect scores ranged from $-0.675$ for questions where image presence had negative impact, all the way to $0.75$ for those questions where images have a positive impact on accuracy. Notably, the negative image effect was for a relatively small proportion of image questions ($25\%$). We noticed that when the original article image is of poor quality~\footnote{To evaluate image quality, we checked whether images fit the image quality criteria from the manual of style on Wikipedia\cite{manualofstyle}. These include basic photographic quality rules, such as exposure and clutter, as well as more generic rules to optimally represent the article’s subject.} (see Figure~\ref{fig:imagequestion}(a)), participants struggled with image recognition questions. Also, we noticed lower accuracy when article images depicted the subject as a whole, while image recognition questions depicted only a part or detail of the subject, or vice versa (see Figure~\ref{fig:imagequestion}(b)), or when image recognition questions depicted the subject in form of illustration, while the article image was a photographic rendering of the subject (see Figure~\ref{fig:imagequestion}(c)). Image presence had no significant impact on image recognition question accuracy when the subject of the article was inherently difficult to distinguish visually, for example, for the article about \textit{Victoria Falls}. Finally, participants who saw illustrated articles failed when image recognition questions were trickier than average. Harder questions included: \textit{``Which of these is a \textit{male} upland goose?''}, or\textit{ ``Which of these is \textit{not} the Taj Mahal?''}

\section{Discussion and Conclusions}

Wikipedia is enormously important for helping people learn about a variety of topics. Images from Wikipedia are an integral part of this experience, both on Wikipedia and where the content is repurposed (e.g., search engine result pages). Despite the many ``views'' Wikipedia images get every day, we know little of the role that images play in people's information uptake. In this work, we contribute an open-source curated question \& answer dataset to test knowledge comprehension and recall in Wikipedia. Additionally, our study  makes several empirical contributions:

\paragraph{Learning about general knowledge is rarely supported by images.}  We found that images do not always help learning about general knowledge facts of a topic---in fact, participants were as likely to correctly answer general knowledge questions when the Wikipedia article did not include an image as they were when there was an image. However, our experiment revealed specific cases where images \textit{did} support answering general knowledge questions (see Fig. \ref{fig:nonimagequesiton}(a)). For example, remembering when a building was built or a painting was created may be easier with an image if the viewer can make inferences based on how that building or painting looked.  If a viewer generally recalls the properties of a Romantic painting, they may be able to accurately estimate when a painting was made (even without seeing the date in text). For some topics (and with some background knowledge), images on Wikipedia may help readers create a mental model of an object that they can then use for inferences~\cite{gyselinck1999role}. Future work may more generally identify when such images help and why.

\paragraph{Learning about visual knowledge is only supported by images if they are consistent with the text.} Interestingly, the presence of images did not significantly improve answer accuracy for visual knowledge questions, such as questions about materials, color, or size. In fact, having an image or not resulted in the same answer accuracy in our study, suggesting that images, in general, do not support visual knowledge learning. This is perhaps the most surprising result of our study, as images are commonly assumed to aid engagement and learning~\cite{clark2010graphics}. However, this result is not entirely inconsistent with past work. Related prior studies on image use in more general online contexts has also found that images may not support content retention~\cite{beymer2007eye}. 

In our follow-up analysis, we found that images that are consistent with their visual knowledge questions are answered accurately significantly more often than images that are inconsistent with the question.  Images seem to boost answer accuracy when they are closely aligned with the text. This is in line with prior work on misaligned or `seductive' images
(e.g.,  \cite{harp1997role,peeck1993increasing}). Guides for teachers recognize this as the first of the 10 recommendations for pictorial facilitation of text information  (the ``10 tenents for teachers'') from Carney and Levin \cite{carney2002pictorial}:  ``Select pictures that overlap with text content''. However, in our visual consistency annotations, we found that less than 40\% of the image-visual knowledge question pairs in our dataset are aligned. Since visual knowledge questions are generated to reflect the main (and potentially more interesting) features described in the article text, this  suggests that many images in Wikipedia  are not optimally chosen to reflect the textual content in the article. While not all visual characteristics are equally important, depending on the learning context, our initial experiment suggests a number of hypotheses for future study. For example: that Wikipedia editors may find it difficult to anticipate what makes a good image; that no single image can illustrate all important characteristics; or that the Wikimedia Commons does have great images choices to pick from.

\paragraph{Images help visual recognition.}
We found that images generally enhance the ability of readers to visually recognize other instances of the object described in an article, compared to relying only on textual information. 
These findings suggest that images can be quite useful in encyclopedic settings, since the ability to visually recognize and classify objects is a fundamental aspect of everyday life \cite{bruce1994recognizing}. Unillustrated articles, while providing enough useful information to learn about general aspects of a subject, do not offer a full visual description. This may naturally affect the reader's ability to recognize objects of that class or instance. In fact, what determines the accuracy of our image recognition questions is not whether the participant read the article (image recognition question scores are the same for the baseline and experiment conditions), but whether an image was present when the articles were read. 

These results are consistent with theories of images helping with retention. For example, dual coding~\cite{paivio1990mental} has been applied in educational contexts to justify the use of images~\cite{clark2010graphics,vekiri02} and identify their potential limitations~\cite{harp1997role}. While there is evidence in our results that Wikipedia images may not be optimally chosen, it is important for future work to reflect on what we are optimizing for. Our choice of visual knowledge questions was driven by the specific article text---that is, aspects of category or instance that were judged important enough to be included in the article. While these were not `outlier' questions (e.g., one in one million birds is red instead of blue), we did not rank the facts/questions based on importance. 

The idea of `importance' connects to various theories of categories. For example, we could consider the classic linguistic notion of necessary and sufficient properties by which we would classify an object into a category. These theories posit that it is possible to create categorical boundaries and hierarchies through this classification. Those properties (e.g., the unique shape of a bird's wing and coloring) may be the most crucial to emphasize both in text and in an image. However, it is not clear that all items in Wikipedia could be reduced to this model. Wikipedia, and our dataset, reflect pages for both classes (e.g., Totem Poles) and instances (the painting `La Scapigliata'). For the latter, it is not obvious that classifying an item, such as a painting, by either things that make it unique or things that put it in a category, is appropriate. We may simultaneously want to understand what makes an item look unique and look similar or different to other related items (e.g., other paintings from the time period). This makes the selection of both `key facts' and good image challenging. Prototype theory~\cite{rosch1973natural} reflects an alternative framework, though largely a linguistic one, in that one can judge how close categories are to each other. For example, one could argue that a robin is a better prototype of a bird than is a penguin (based on general understanding). By extension, the characteristics of a robin are a better representation of the bird category. Insofar as we can extend this to visual properties, one could ask `how much would you associate a visual fact (e.g., the blue coloring) with the category or instance?' We note that while we did not explicitly seek prototypical descriptions when crafting questions, we did avoid those that might only apply to outliers. However, it should be emphasized that prototypical visual characteristics may not be the most important ones for a learning task. For example, we might care about non-prototypical characteristics when describing the similarity of two things. As an area of future work, we might use domain experts to re-rank facts, and therefore questions, by how `important' they are to the teaching of a concept. This would allow for a more refined analysis of the (in)consistent image question rather than a simple binary one. Thus, we could answer not just the question, `does this image express a particular characteristic or set of characteristics from the text?' but also `how well does this image meet our specific learning objectives?'

\paragraph{Photographic quality matters.} The cognitive theory of multimedia learning suggests that higher quality visuals reinforce and extend knowledge \cite{mayer2002multimedia}.  Upon manual inspection, we found that there might be some relation between photographic quality and knowledge acquisition from Wikipedia articles. We found that images of low photographic quality (see, for example,  Figure~\ref{fig:imagequestion}(a)) can hinder participants' ability to visually recognize the subject of the article. Guidelines for image use in field guides--books used to recognize and research wildlife--suggest that high-quality images are essential to make good educational material, especially when it comes to introductory sections and glossaries \cite{plantguide}. We also found that people may find it difficult to generalize  the insight from an image to other instances. For example, when shown an illustrated version of a plant, our participants were unable to accurately recognize the same plant in a photographed version. This could indicate that a dual representation is helpful, especially for article topics that are sought for learning what something looks like. It also indicates that Wikipedia images are not always chosen to include representative features to be perceived as being of the same category.

\subsection{Design Implications}
Collectively, our findings have several theoretical and practical implications.

\paragraph{Promote the addition of missing images on Wikipedia} 
Wikiwork---human contributions that improve Wikipedia in various ways---is extremely varied. Some editors add new articles, while others focus on improving citations or correcting grammar mistakes. Unfortunately, it is not obvious that image-focused wikiwork is as widely understood or appreciated. 
We see a reflection of this in the Barnstars awarded by the Wikipedia community to editors~\cite{kriplean,barnstars}. Barnstars can be awarded for everything from diligence to being a `surreal contributor' (``any Wikipedian who adds `special flavor' to the community by acting as a sort of wildcard''). The range of these awards provide a partial signal of what kind of work is valued. While there are hundreds of types of Barnstars only a handful are dedicated to any kind of image work.
For example, an SVG Barnstar is awarded to those converting raster to vector images. A Valued Picture Barnstar is awarded to, ``users that uploaded a free image to Wikipedia that is considered wonderful, valuable, and/or crucial'' and the Photographer Barnstar goes to users ``who tirelessly improve Wikipedia with their photographic skills and contributions.''~\cite{barnstars}.

Our work demonstrates the importance of good images for a variety of learning objectives. As learning represents a key use case for Wikipedia~\cite{singer2017we}, wikiwork focused on curating, finding, or creating good images should be more valued. Although Barnstars alone are unlikely to provide motivation on their own, acknowledging work is a key component in increasing editor retention and norm-setting~\cite{matiasvolunteers}. Exactly what this kind of image-focused contribution looks like, and what types should be prioritized, is an extensive topic for research. However, these measures might be included within community-led efforts to add images to unillustrated Wikipedia articles, such as the already existing \textit{Wikipedia Pages Wanting Photos \#WPWP} campaign \cite{WPWP}, or the \textit{Visible Wiki Women} \cite{visible} initiative, as well as those ``Wiki Loves'' campaigns designed to improve the visual coverage of encyclopedic topics, such as monuments (\textit{Wiki Loves Monuments} \cite{WLM}), or natural heritage (\textit{Wiki Loves Earth} \cite{WLE}). In addition to `novel' types of wikiwork, image-centered contributions can also become integrated into existing structures. For example, editors within a subtopic (e.g., birds, sociology, food, etc.) will often agree on templates and structures for those pages around text or infoboxes. Such constraints are difficult for images. For example, we may want all bird pages to have a consistent look and feel for their images, but the perfect image may not exist (or may not be in fair-use). While subject-matter experts disagree over whether photographs or paintings/illustrations are best for characterizing the world~\cite{birdguide,plantguide}, there is general agreement that a consistent look is important.
This presents a unique and interesting sociotechnical challenge. The best interventions to normalize and prioritize image-centered wikiwork, all within the unique `bureaucracy' of Wikipedia~\cite{butler2008don}, present important future topics for research. 

\paragraph{Promote the addition of missing images to Wikidata}
The findings in our paper also highlight the importance of adding high-quality images to Wikidata. Wikidata is a Wikimedia project and the open, manually curated knowledge graph that allows one to connect all Wikipedia articles across different languages. For example, the Train article exists in 162 languages, that often inherit structured information and the infobox image from the corresponding Wikidata item \cite{wikdiatatrain}, either manually or automatically.  
Such an image may not be `best' or even `good' and may not be culturally appropriate. While highly edited languages may counteract this kind of behavior and replace the image, languages with lower editorial resources may not. 

\paragraph{Encourage the addition of images about historic eras} Our findings suggest that articles that describe aspects of historical time periods, such as specific paintings or architectural artifacts of a certain time, may benefit from accompanying images. While architecture and visual arts figure among the top topics by image presence on Wikipedia \cite{gapsmap}, a large percentage of these articles still remain unillustrated. A concrete step building on our results could be to promote initiatives that encourage articles about such topics include high-quality images. Moreover, existing editor groups who work on improving Wikimedia content on these topics, such as WikiProject Cultural Heritage on Wikidata \cite{wdculturalheritage}, or Wikiproject Historic Sites on Wikipedia \cite{wphistoricareas}, could explicitly encourage contributions to the visual content of the articles and items in these domains.

\paragraph{Encourage or suggest images that are consistent with the text.} While Wikipedia editors can freely choose what images to add, our findings show that the chosen images are not always optimal. In fact, our analysis showed that images that are relevant to the text lead to higher accuracy in answering visual knowledge questions. Wikipedia's manual of style \cite{manualofstyle} already suggests to choose consistent images, but our dataset shows that this is not always followed.  While the manual of style speaks to the importance of image relevance to the generic ``topic's context'', we find here that images that are not aligned with the visual description of the subject can hinder the ability to learn. Additional guidelines and examples may be needed for ensuring that images are consistent with descriptive text and to alert editors of the negative impact of inconsistent images.  Additionally, existing multimodal machine learning frameworks that measure the alignment between images and text \cite{radford2021learning,srinivasan2021wit}, could be incorporated into image suggestion frameworks to select the best images for unillustrated articles, such as the \textit{add-an-image} task for newcomer editors \cite{addimage}.

\paragraph{Discourage low-quality images.} Our study also reveals that images of low quality can hinder participants' ability to visually recognize the subject of the article.  The English Wikipedia manual of style \cite{manualofstyle} indeed suggests that images of ``Poor-quality images—dark or blurry; showing the subject too small, hidden in clutter, or ambiguous; and so on—should not be used unless absolutely necessary''. Automated image quality classifiers \cite{beytia2022visual} have been trained based on images labeled as ``Quality Images'' \cite{qualityimages} by the Wikimedia Commons community.  
Given resource limitations on good fair-use images, creating images may be crucial. Selecting culturally appropriate images for different Wiki language domains may also be an important form of labor. To promote the usage of high quality images, similar computer vision-based classifiers could be incorporated into image suggestion algorithms, or search tools, that help editors find the right images for Wikipedia articles.  

\subsection{Limitations and Future Work}
We hope that our manually curated question \& answer dataset to test knowledge comprehension in English Wikipedia can be used and extended for other studies of images on Wikipedia. For example, our current dataset is focused on showing one specific image alongside the text (the so-called `lead' image). However, there is no evidence that this is the best image of a concept that Wikipedia or the Commons has to offer. A natural extension would be to experimentally vary the image displayed alongside the text with others that are available in the Commons. Such a use would provide guidance for editors, and potentially tool builders, in identifying not just a reasonable image, but the best one among a set.

A limitation of our dataset is that our questions were manually generated by a small group of researchers. Although we asked the annotators to follow a thorough process and take into account the interestingness of the questions for the general public, the pool of questions in our data might not reflect the actual information needs and interests of Wikipedia readers. Future research is needed to align our reading comprehension questions with readers' motivations, and to analyze how images are useful to satisfy the different information needs.

Moreover, while the dataset is the first of its kind, its scale can be further expanded by automating article, section, image, and question retrieval using Natural Language Processing to detect Wikipedia articles which include a physical description and automatically generate questions that can be verified in crowdsourcing settings.  Such larger datasets would enable larger-scale experiments, which could make use of automated analysis of visual content to determine which subjects, spatial arrangements, and image types are more useful to improve reading comprehension.

Alignment between images and their corresponding captions also plays an important role in text comprehension \cite{bernard1990using}. This topic even has a dedicated page in Wikipedia's manual of style \cite{manualofstyle}. While investigating the importance of captions in Wikipedia knowledge acquisition was outside the scope of this work, this is an area for future research which could be implemented with the data and framework provided with this study.

Our study is also limited by being offered only in English. Our findings may not generalize to other languages and people from other countries and cultures. Future work could extend our study to a more diverse population to study the helpfulness of images with a sample more representative of Wikipedia users. 

In this initial work, we have focused on both a set of assumed motivations and learning objectives for reading Wikipedia. Although we believe that our choices are a good proxy for common tasks on Wikipedia, they do not represent the full breadth and complexity of the space. Additional research, with more types of images, articles, levels of expertise, learning objectives, assessments, and so on, will further clarify which images are contextually useful.

\section{Conclusion}

In this work, we set out to evaluate whether images on Wikipedia support a set of common learning tasks. Our results build on research to understand the impact of text and media on learning. Wikipedia is a particularly important platform to analyze due to its broad impact and unique crowdsourced organization. 

Using a novel dataset of article excerpts, images, and questions we conducted a large-scale online experiment (n=704) with participants on LabintheWild and Prolific. Our results reveal the specific utility of images on different types of questions. Encouragingly, we did not find situations where images markedly harm learning tasks. For image recognition tasks, article images significantly improved reader accuracy in selecting target concepts (e.g., \emph{which of these pictures is a leafy seadragon?}). We also found that for visual knowledge questions, only a subset of images help. Given the immense reach of Wikipedia images, our results have broad implications for policy, design guidelines, and novel tools.
\section*{Acknowledgements}
We are grateful to the NSF for their support of this work through NSF IIS-1815760 and NSF IIS-2006104.

\section*{Image Attributions}
\textbf{Figure 1:}
\begin{itemize}[wide = 0pt]
\item 
\textit{Leafy Seadragon on Kangaroo Island.jpg}; James Rosindell, CC BY-SA 4.0
\end{itemize}

\noindent \textbf{Figure 2: }
\begin{itemize}[wide = 0pt]
\item \textit{2005-09-20 1080x1920 chicago 900 north michigan.jpg;} J. Crocker, Attribution
\end{itemize}

\noindent \textbf{Figure 5 (c):}
\begin{itemize}[wide = 0pt]
\item \textit{Gujia dry.JPG}; Prasant iitd, CC BY-SA 4.0
\item \textit{Fried dumpling (3337901416).jpg}; pelican from Tokyo, Japan, CC BY-SA 2.0
\item \textit{Xia jian jiao.jpg}; lazy fri13th, CC BY-SA 2.0
\item \textit{Deep Fried Cholent Dumpling (F\"ulem\"ule Restaurant)}\\ \textit{(41496749174).jpg}; Asok5 from Budapest, Hungary, CC BY-SA 2.0 
\end{itemize}

\noindent \textbf{Figure 8 (a):}
\begin{itemize}[wide = 0pt]
\item \textit{Irelands history.jpg}; Tjp finn, CC BY-SA 4.0
\item \textit{Wwp50roof.JPG}; Jim.henderson at the English Wikipedia, Public domain
\item \textit{Leonardo da vinci - La scapigliata.jpg}; Leonardo da Vinci, Public domain
\end{itemize}

\noindent \textbf{Figure 8 (b):}
\begin{itemize}[wide = 0pt]
\item \textit{Florenz - Bargello 2014-08-09r.jpg}; Rabe!, CC BY-SA 4.0
\item \textit{Two Brothers totem pole Jasper.jpg}; Ymblanter, CC BY-SA 4.0
\end{itemize}

\noindent \textbf{Figure 10 (a):}
\begin{itemize}[wide = 0pt]
\item \textit{Spectral bat photo.jpg}; Marco Tschapka, CC BY-SA 3.0
\item \textit{Leafy Seadragon on Kangaroo Island.jpg}; James Rosindell, CC BY-SA 4.0
\end{itemize}

\noindent \textbf{Figure 10 (b):}
\begin{itemize}[wide = 0pt]
\item \textit{2005-09-20 1080x1920 chicago 900 north michigan.jpg}; J. Crocker, Attribution
C\item \textit{hicago 900 North Michigan Avenue (2827940356553).jpg}; Roman Boed from The Netherlands, CC BY 2.0
\end{itemize}

\noindent \textbf{Figure 10 (c):}
\begin{itemize}[wide = 0pt]
\item \textit{Calamintha nepeta Sturm54.jpg}; Johann Georg Sturm (Painter: Jacob Sturm), Public domain
\item \textit{Catnip-blossom.jpg}; Jon Sullivan, Public domain
\end{itemize}

\balance

\bibliographystyle{ACM-Reference-Format}
\bibliography{bibliography}

\end{document}